\documentstyle[twocolumn,psfig,prb,aps]{revtex}

\begin{document}

\draft
\tighten

\title
{{\it Ab initio} calculation of the potential energy surface\\
for the dissociation of H$_2$ on the sulfur covered Pd(100) surface\\}
 
\author{C. M. Wei\cite{Add}, A. Gro{\ss}, and M. Scheffler}
\address
{ Fritz-Haber-Institut der Max-Planck-Gesellschaft,\\ 
Faradayweg 4-6, D-14 195 Berlin-Dahlem, Germany }
 
\maketitle
 
\begin{abstract}
The presence of sulfur atoms on the Pd(100) surface is known to hinder the 
dissociative adsorption of hydrogen.  Using density-functional theory and the 
full-potential linear augmented plane-wave method, we investigate the 
potential energy surface (PES) of the dissociative adsorption of 
H$_{2}$ on the sulfur covered Pd(100) surface. 
The PES is changed significantly compared to the dissociation on the clean 
Pd(100) surface, in particular for hydrogen close to the S atoms.
While the hydrogen dissociation at the clean Pd(100) surface is non-activated,
for the (2$\times$2) sulfur adlayer (coverage $\Theta_{\rm S}$= 0.25) 
the dissociation of H$_2$ is inhibited
by energy barriers. Their heights strongly depend on the distance 
between the hydrogen and sulfur atoms leading to a highly corrugated PES.
The largest barriers are in the vicinity of the sulfur atoms due to the
strong repulsion between sulfur and hydrogen.
Still the hydrogen dissociation on the (2$\times$2) sulfur covered Pd(100) 
surface is exothermic. Thus the poisoning effect of sulfur adatoms for H$_{2}$ 
dissociation at low sulfur coverage ($\Theta_{\rm S}$ $\leq$ 0.25) is mainly 
governed by the formation of energy barriers, not by blocking of the 
adsorption sites.  
For the c(2$\times$2) sulfur adlayer ($\Theta_{\rm S}$= 0.5), 
the PES for hydrogen dissociation is purely repulsive. 
This is due to the fact that for all different 
possible adsorption geometries the hydrogen molecules
come too close to the sulfur adatoms before the dissociation
is completed.
\end{abstract}
\pacs{PACS numbers: 68.45.Da, 73.20.At, 82.65.Jv}
 
\section{Introduction}
 
The modification of the chemical reactivity and selectivity of
surfaces by adsorbates is important for, e.g., building better catalysts.
Therefore the investigation of the microscopic mechanisms of how adatoms
poison or promote certain reactions is -- besides its fundamental interest --
of great technological relevance. Hydrogen dissociation has become 
the benchmark system for theoretical and experimental
studies of simple chemical reactions on surfaces. 
On clean metal surfaces, a detailed picture of how H-H bonds are broken 
and how new bonds between the hydrogen atoms and the surface are formed 
has been developed by theoretical 
studies.\cite{Lun79,Nor81,Fei91,Ham94,Whi94,Ham95PRL,Whi96,Wil96PRB,Eich96} 
The modification of the potential energy surface 
(PES) of hydrogen dissociation on an adlayer-covered surface has not 
been addressed in similiar detail by theory so far.

Experimentally it is well established that the presence of sulfur
on metal surfaces causes a drastic reduction in hydrogen sticking 
probabilities.\cite{Good81,John81,Ren89,Bur90}
Earlier theoretical studies explaining the poisoning mechanism
focused on general concepts. Feibelman and Hamann
\cite{Fei84,Fei85} suggested that the poisoning effect of sulfur is related 
to the sulfur induced change of the density of states (DOS) at the Fermi level.
This explanation was also employed by MacLaren and co-workers.\cite{MacL86} 
Current studies, however, have emphasized that the reactivity of surfaces
cannot be solely understood by the DOS at the Fermi level.\cite{Ham95,WCoh96}
A different model for the poisoning was proposed by N{\o}rskov
{\it et al.}.\cite{Nor84,Nor93,Ham93} These authors explain the modification 
of the reactivity by adlayers by the interaction of the H$_2$ 
molecule with the adlayer induced electrostatic field. 
These models, however, do not provide a detailed microscopic picture of 
how the hydrogen dissociation process is actually modified by the adlayer, 
i.e., either by blocking adsorption sites for atomic hydrogen 
or by building up energy barriers along the dissociation pathways of H$_2$.

\vspace{-11.2cm}

\hspace*{.1cm} {\large {\tt subm.~to~Phys.~Rev.~B,~Nov~1997}}

\vspace{10.8cm}

Recently the poisoning mechanism of Pd(100) by a sulfur adlayer has been 
investigated theoretically by Wilke and Scheffler.\cite{Wil95,Wil96}
The Pd(100) surface is a well-suited substrat for the investigation 
of the poisoning mechanism of catalytic reactions,
because many experimental and theoretical studies exist and form
a wealth of information for comparison. At the clean surface, hydrogen
molecules dissociate spontaneously, i.e., non-activated 
dissociation pathways exist with no hampering energy 
barrier.\cite{Wil96PRB,Ren89,Bur90,Com80,Behm80}
When the surface is covered with sulfur, the H$_2$ sticking probability 
is significantly reduced.\cite{Ren89,Bur90} 
These experiments show that with increasing
sulfur coverages $\Theta_{\rm S}$ the initial sticking coefficient of H$_2$ 
strongly decreases. This is true in particular for molecules
with low kinetic energy \cite{Ren89} ($\leq$ 0.1 eV) that can adsorb only if 
nonactivated dissociation pathways exist and are accessible. For 
$\Theta_{\rm S} \approx$~0.25 the initial sticking coefficient of molecules 
with low energies is approximately two orders of magnitude smaller than the one
for the clean surface,\cite{Ren89,Bur90} indicating that for this 
sulfur coverage
non-activated dissociation is nearly completely hindered. When the
sulfur coverage is increased, the initial sticking coefficient reduces
even further; at $\Theta_{\rm S} \approx$~0.5 it is about three orders 
of smaller than at the clean surface.\cite{Ren89}
In addition to the significant decrease of the initial sticking coefficient,
TPD studies \cite{Bur90} observed a decrease of the hydrogen saturation 
coverage with increasing $\Theta_{\rm S}$. Burke and Madix \cite{Bur90} 
therefore concluded that sulfur adatoms substantially reduce 
the hydrogen adsorption energy at sites in their vicinity, making these 
positions unstable against associative desorption. An earlier permeation 
study of Comsa, David, and Schumacher \cite{Com80} led to the
same conclusion. 

These findings are at variance with the results of Wilke and 
Scheffler.\cite{Wil95,Wil96} They showed by density-functional theory
calculations that the hydrogen dissociation
at the (2$\times$2) sulfur covered Pd(100) surface is still exothermic
and that the blocking of adsorption sites is therefore only of minor
importance at low coverages. Adsorbed sulfur builds up energy barriers in the 
entrance channel and thus hinders the dissociation. Their results gave a clear 
explanation for the poisoning mechanism. We have now extended their work.
We have investigated in detail how the poisoning of the hydrogen dissociation
on Pd(100) by the presence of sulfur depends on the position and orientation
of the molecule. Furthermore, we have addressed the influence of the sulfur
coverage on the poisoning.

In this paper, using density-functional theory and the full-potential linear
augmented plane wave (FLAPW) method, we calculate the PES of 
H$_2$ dissociative adsorption over the (2$\times$2) and the c(2$\times$2) 
sulfur covered Pd(100) surface at different adsorption sites and different 
orientations, and thus provide a complete picture of the poisoning mechanism 
caused by the adsorbed sulfur on Pd(100) surface. We fully confirm the 
conclusions of Wilke and Scheffler\cite{Wil95,Wil96}: the poisoning effect 
of sulfur adatoms for H$_{2}$ dissociation over Pd(100) surface at low 
sulfur coverage ($\Theta_{\rm S}$ $\leq$ 0.25) is governed 
by the formation of energy barriers. The height of the energy barriers
depends strongly on the distance between H$_{2}$ and the S atoms. For high 
sulfur coverage ($\Theta_{\rm S}$= 0.5), the PES for hydrogen dissociation 
becomes purely repulsive due to the fact that for all possible different 
adsorption geometries the hydrogen molecules come too close to the sulfur 
adatoms before the dissociation is completed. The results are now sufficiently 
complete for a six-dimensional quantum dynamical simulation \cite{Gro95PRL}.
This requires the representation of the {\it ab initio} PES by an
analytical function \cite{Gro97PRB} in order to interpolate the actually
calculated points and to achieve a closed expression of the PES. This
will be also presented in this paper.

The structure of the paper is as follows. In Section II we describe 
the theoretical method and computational details. The potential energy
surface for the hydrogen dissociation on the (2$\times$2) sulfur covered
Pd(100) surface at different adsorption sites is presented and discussed in 
Section~III. Section IV reports the results of the H$_2$ interaction with the 
c(2$\times$2) sulfur covered Pd(100) surface. The dependence of the PES on the 
orientation of the molecule is analysed in Section V. After a brief description
of the analytical representation of the PES suitable for a quantum dynamical 
simulation in Section VI, the paper concludes with a summary in Section VII.

\section{Computational details}
       We have used density-functional theory 
together with the generalized gradient approximation (GGA) \cite{Per92} 
for the exchange-correlation functional.  The full-potential linear 
augmented plane wave (FLAPW) method \cite{Bla95,Koh96} is employed for 
solving the non-relativistic Kohn-Sham equations.  The FLAPW wave 
functions in the interstitial region are represented using a plane 
wave expansion up to $E_{\rm cut}$ = 11 Ry.  For the potential representation 
plane waves up to $E_{\rm cut}$ = 169 Ry are taken into account due to 
a small muffin-tin radius around the H atoms (r$^H_{MT}$= 0.37 \AA).  
Inside the muffin-tin spheres the wave functions are expanded in 
spherical harmonics with $l_{max}$ = 10, and non-spherical 
components of the density and potential are included up to $l_{max}$ = 4.  
For the {\bf k}-space integration, a grid of 5$\times$5 
(or 6$\times$6) uniformly 
spaced points are used in the two-dimensional Brillouin zone of 
the (2$\times$2) [or the c(2$\times$2)] surface unit cell, 
and we find that for a finer mesh
the adsorption energies change by at most 30 meV for the (2$\times$2) hollow 
site adsorption geometry.  

We used a supercell geometry and modeled the metal surface by three-layer 
slabs which were separated by a 12 \AA\ vacuum region.  
Hydrogen and sulfur atoms were placed at both sides of the slab.  
For the geometry of the sulfur covered Pd(100) surface, the results 
reported in Ref.~\onlinecite{Wil95}  were used. 
Tests were made for a five-layer slab at different adsorption 
geometries, and the average total energy difference from a three-layer slab 
calculation was found to be less than 40 meV (see Table~\ref{tab1}). 
This agrees with results reported by Wilke and Scheffler~\cite{Wil96}
who also studied the influence of the slab thickness on the potential
energy for slabs with up to seven layers. They found energy differences of
less than 30~meV for the barrier heights and the adsorption energies
at the hollow sites of the surface.

  The substrate geometry was kept fixed for the H$_{2}$ adsorption 
pathways studied.  This is appropriate and plausible because the mass 
mismatch of H with the substrate atoms is significant.  The energy zero is 
taken as the energy of the geometry where the molecule is sufficiently far 
away from the surface ($Z$ = 4.03 \AA) such that there is practically 
no interaction between the molecule and the surface. Zero-point 
corrections are not included in the PES.

\section{H$_{2}$ at the (2$\times$2) sulfur covered P\lowercase{d}(100)
surface}
\subsection{Determination of the potential energy surface (PES)}
Neglecting surface relaxation effects, the potential energy surface for the 
dissociative adsorption of a hydrogen molecule over a sulfur covered Pd(100) 
surface is six-dimensional corresponding to the molecular degrees of freedom.
The coordinates used in this work are defined in Fig.~1. 
The H$_2$ center of mass position is given by three Cartesian 
coordinates ($ X, Y, Z$) for which the origin is 
chosen as the hollow position of the topmost Pd layer. The two 
rotational degrees of freedom are given by the angle {\bf $\theta$} of 
the molecular axis with the $Z$-axis (cartwheel rotation) and 
the angle $\phi$ with the $X$-axis (helicopter rotation). 
The distance between the hydrogen atoms is denoted by $d_{\rm H-H}$.

To map this high-dimensional energy surface the common strategy is to compute 
two-dimensional cuts through the energy surface, so-called elbow-plots, where 
($ X, Y, \theta, \phi$) are fixed, and 
only $d_{\rm H-H}$ (the bond length of the hydrogen molecule) 
and $Z$ (its height above the surface) are varied.
Figure 2 shows the surface unit cell for the (2$\times$2) 
sulfur covered Pd(100) surface. 
Inside the unit cell, there are two bridge sites {\bf b$_1$} and 
{\bf b$_2$}, two hollow sites {\bf h$_1$} and {\bf h$_2$}, the 
top site above the Pd atom {\bf t$_{\rm Pd}$}, and the top site above the S 
atom {\bf t$_{\rm S}$}. In order to obtain a comprehensive information about 
the adsorption behaviorof H$_2$, we have evaluated the elbow-plots at 
different cuts through the six-dimensional configuration space of H$_2$ 
where the cuts are defined by the site $(X, Y)$ and the molecular 
orientation ($\theta$, $\phi$).

\subsection{Results}
To distinguish between the different molecular geometries, we characterize 
them by the set of fixed coordinates ($X, Y, \theta, \phi$).  
For example, if one defines {\large $a$} as the unit length 
of the (2$\times$2) unit cell, 
then (0.5{\large $a$}, 0.5{\large $a$}, 0$^\circ$, 0$^\circ$) refers 
to a geometry with the hydrogen molecule placed at the {\bf h$_1$} 
site in an upright position (molecular axis perpendicular to the surface); 
and (0.5{\large $a$}, 0.5{\large $a$}, 90$^\circ$, {\bf $\phi$}) refers 
to a geometry with the molecule at the {\bf h$_1$} site with its  
axis parallel to surface.  The geometry in which the molecule is parallel 
to the surface can also be characterized by the positions where the 
hydrogen atoms will be adsorbed, and by that over which the center of mass 
of the molecule is situated.  For example (0.5{\large $a$}, 0.25{\large $a$}, 
90$^\circ$, 90$^\circ$) can be denoted by {\bf h$_1$-b$_1$-h$_2$} which 
refers to a geometry with the center of mass over the bridge ({\bf b$_1$}) 
position and the atoms oriented towards the two hollow sites {\bf h$_1$} and 
{\bf h$_2$} in which they will finally be adsorbed.

\subsubsection{2-D cuts through the PES at hollow sites}
For the PES of the hydrogen dissociation over the 
(2$\times$2) S covered Pd(100) surface
($\Theta_{\rm S}$= 0.25), two adsorption geometries {\bf b$_1$-h$_1$-b$_1$} 
and {\bf b$_1$-h$_2$-b$_1$} have been reported by Wilke and 
Scheffler.\cite{Wil96}  Their results show that 
the presence of a the (2$\times$2) S 
adlayer on Pd(100) changes the dissociation process significantly from 
being non-activated at the clean Pd(100) surface to activated 
dissociation. In their calculations, the dissociation pathways corresponding 
to the {\bf b$_1$-h$_1$-b$_1$} and {\bf b$_1$-h$_2$-b$_1$} 
geometries at the (2$\times$2) sulfur covered Pd(100) surface had energy 
barriers in the entrance channel of 0.1 and 0.6 eV, respectively.  

Figure 3(a) shows the re-calculated PES for the {\bf b$_1$-h$_1$-b$_1$} 
adsorption geometry. The dissociation pathway 
has an energy barrier of 0.11 eV~in the 
entrance channel which is consistent with previous calculations.\cite{Wil96} 
This configuration corresponds to the minumum barrier dissociation 
pathway. We will show below that the energy barriers increase
as a function of decreasing distance of the hydrogen molecule to the sulfur 
atoms on the surface. For that reason it is not surprising that the 
{\bf b$_1$-h$_1$-b$_1$} adsorption pathway exhibits the lowest energy
barrier because the {\bf h$_1$} hollow site is the site on the surface
furthest away from the sulfur atoms. 

Figure 3(b) shows the PES of the {\bf t$_{\rm Pd}$-h$_1$-t$_{\rm Pd}$} 
pathway. At the clean Pd(100) surface, this path has a 
local minimum potential energy of -0.23 eV.\cite{Wil97}
The calculated results show that upon the adsorption 
of a (2$\times$2) S adlayer an activation energy barrier 
towards dissociative adsorption of 0.13 eV builds up
in the entrance channel at ($Z$, d$_{H-H}$) = (1.7 \AA, 0.78 \AA), 
and the minimum at ($Z$, d$_{H-H}$) = (0.98 \AA, 0.94 \AA)
becomes very shallow with an energy of 0.01 eV.
For this geometry, the two hydrogen atoms  
are about 3.3 \AA\ away from the adsorbed sulfur atoms, hence
one can expect that there is no direct interaction between H and S atoms.
In such a situation the energy increase is due to the sulfur-induced
modification of the local electronic structure at the surface Pd atom
\cite{Wil96}, as will be confirmed below.

Figure 3(c) shows the PES for the {\bf b$_2$-h$_2$-b$_2$} pathway. 
In this geometry the dissociation process becomes purely repulsive; it
differs from the behavior at the {\bf b$_1$-h$_2$-b$_1$} geometry where 
activated dissociation with an energy barrier of 0.6 eV is
possible.\cite{Wil96}
This is easy to understand because the hydrogen molecule dissociates 
towards the sulfur atoms at the {\bf b$_2$-h$_2$-b$_2$} geometry, 
and the direct interaction between H and S atoms causes 
the dissociation process to become purely repulsive. 

        Figure 3(d) shows the PES for a geometry with the hydrogen molecule 
adsorbed at the {\bf h$_1$} site in an upright position,
i.e. with its molecular axis perpendicular to the surface.  
At the clean surface, there exists a shallow local minimum of -0.03 eV at
($Z$, $d_{\rm H-H}$)= (0.60 \AA, 0.96 \AA).\cite{Wil97}
As is evident from Fig. 3(c), this pathway becomes repulsive 
upon the adsorption of the (2$\times$2) S adlayer and the energy at 
($Z$, $d_{\rm H-H}$)= (0.60\AA, 0.96 \AA) has raised to 
0.28 eV from -0.03 eV. There is still a shallow well, however.
 For this geometry, the two hydrogen atoms 
directly above the {\bf h$_1$} site are at least 4 \AA\ away 
from the adsorbed sulfur atoms, so there is no direct interaction 
between H and S atoms. Thus the increase in the potential energy 
has to be due to the sulfur-induced modification of the local electronic 
structure at this site.

\subsubsection{2-D cuts through the PES at bridge sites}
  In addition to the dissociation over the {\bf h$_1$} and {\bf h$_2$} 
hollow sites, we also considered geometries where the center 
of mass of molecule is situated at the bridge site and the top sites 
over the Pd or S atoms.  In Figures 4(a) and 4(b), the PES for the
{\bf h$_1$-b$_1$-h$_2$} and {\bf t$_{\rm Pd}$-b$_1$-t$_{\rm Pd}$} 
pathways are presented.  

At the clean surface, the dissociation pathway for
{\bf h$_1$-b$_1$-h$_2$} is non-activated.
In Fig. 4(a), the PES for {\bf h$_1$-b$_1$-h$_2$} geometry indicates that
the adsorption of H$_{2}$ becomes an activated process with an 
energy barrier of 0.16 eV in the entrance channel 
($Z$=1.8 \AA\  and $d_{\rm H-H}$= 0.78 \AA). 
This energy barrier also results from the modification of the local 
electronic structure.\cite{Wil96}
There exists a second barrier of 0.13 eV
in the exit channel at ($Z$, $d_{\rm H-H}$)=(0.97 \AA, 1.05 \AA). We believe 
that this barrier might be due to the compensation between the direct 
attractive interaction of the Pd atoms with one H atom and the direct 
repulsive interaction of the S atoms with the other H atom, since 
the top of the energy barrier is situated at the value of 
$Z$ close the sulfur adsorption height (1.24 \AA).  

  Figure 4(b) shows the PES for the {\bf t$_{\rm Pd}$-b$_1$-t$_{\rm Pd}$} 
geometry. At the clean surface this dissociation pathway 
has a local energy minimum of -0.2 eV before becoming repulsive.
Our calculations show that, upon adsorption of the S adlayer, 
the dissociation process becomes purely repulsive and is 
different from the behavior at the {\bf h$_1$-b$_1$-h$_2$} geometry
where dissociative adsorption over an energy barrier of 0.16 eV is possible. 
Again, this can be understood by the fact that the hydrogen molecule 
dissociates towards the direction pointing to the Pd atoms and close 
to the S atoms. Thus the direct interaction between the H and S atoms leads
to the disappearance of the shallow minimum along this path 
(E$_{ad}$= 0.20 eV) at the clean Pd(100) surface.

\subsubsection{2-D cuts through the PES at top sites above Pd or S atoms}
The PES for dissociative adsorption over the 
{\bf t$_{\rm Pd}$}, {\bf t$_{\rm S}$} sites near the 
adsorbed sulfur is significantly different from the 
PES at the {\bf h$_1$}, {\bf h$_2$}, and {\bf b$_1$}  
sites that are further away from the adsorbed sulfur.  
Figure 5(a) shows the PES for the {\bf h$_2$-t$_{\rm Pd}$-h$_2$} 
geometry. At the clean surface, this dissociation pathway 
has a local minimum of -0.24 eV in the entrance channel and an energy 
barrier of 0.15 eV in the exit channel [see Fig. 2(c) of Ref. 8].  
The adsorbed sulfur atom is about 2 \AA\  away from the 
adsorption site {\bf t$_{\rm Pd}$} and 1.24 \AA\  above the topmost 
Pd layer.  One can expect that the hydrogen molecule will interact 
directly the S atom before it reaches the 
topmost Pd layer; this therefore raises the energy of the PES. 

        Indeed one can see from Fig. 5(a) that, upon the adsorption of 
the S adlayer, the local minimum of PES in the entrance channel has 
disappeared and the energy barrier in the 
exit channel has raised to 1.28 eV even though the dissociation of hydrogen 
atoms does not point towards the adsorbed sulfur atoms. 

        Figure 5(b) shows the PES at the {\bf t$_{\rm S}$} site within 
{\bf b$_2$-t$_{\rm S}$-b$_2$} geometry. In this geometry, 
the hydrogen molecule directly approaches the S atom on the surface 
which results in a strong repulsion.  
Seen from Fig. 5(b), we found that the PES has increased to more than 
1.5 eV even when the molecule is still 3 \AA\  above the topmost Pd layer.  
However, we found an energy minimum of 1.58 eV at 
($Z$, $d_{\rm H-H}$)= (2.20 \AA, 1.90 \AA).  Judging from the 
fact that the hydrogen atoms are 2.20 \AA\  above the topmost Pd layer, 
we believe that this energy minimum is completely caused by the local 
bonding properties between the H and S atoms.  

        From the facts found in Figure 5, one can expect that if the 
projected distance $d_{||}$ (parallel to surface) 
between the adsorption site and the adsorbed sulfur atom is smaller
than 2 \AA, the dissociation behavior of the
hydrogen molecule is dominated by the strong repulsive interaction 
between H and S atoms. 
This can be further proved by studying the PES of a frozen H$_{2}$ 
molecule (i.e., $d_{\rm H-H}$=const) above the sulfur covered surface.  
Figure 6 shows the PES of a hydrogen molecule with 
$d_{\rm H-H}$= 0.76 \AA\  with its center of mass moving inside the (010) 
plane crossing the {\bf h$_1$}, {\bf t$_{\rm Pd}$}, 
{\bf t$_{\rm S}$} adsorption sites.  
This PES is defined as ({\bf \it X},{\bf \it Y},$Z$,d,{\bf $\theta$}, 
{\bf $\phi$})= ($t/\sqrt(2)$,$t/\sqrt(2)$,$Z$, 
0.76\AA, 90$^\circ$, 135$^\circ$),
where $t$ is the projected distance of the hydrogen center of mass from
the sulfur atom in the (100) plane given in \AA. 

        As one can see from Figure 6, the potential energy of the 
hydrogen molecule increases signficantly when the molecule comes close 
to the sulfur or the palladium atoms.  
The semi-circles in Figure 6 indicate 
the positions where the center-of-mass of the molecule is 1.5 \AA\  away from 
the S atom or 1.2\AA\  away from the Pd atom, respectively.  
The potential energy of the hydrogen molecule at this distance has increased 
to values larger than 1.5 eV. This shows the strong inhibition of the
hydrogen dissociation in the vicinity of the sulfur and palladium atoms.

   Summarizing the results for the H$_2$ dissociation on the
(2$\times$2) S covered Pd(100) surface, over the {\bf h$_1$}, {\bf h$_2$}, 
{\bf b$_1$}, {\bf t$_{\rm Pd}$}, {\bf t$_{\rm S}$} sites we found that the 
dissociation behavior of H$_{2}$ molecule strongly depends on 
the projected distance $d_{||}$ between the hydrogen center of mass
and the adsorbed sulfur atom, and on the orientation of the dissociating
molecule.  The dissociation behavior of hydrogen 
over the the (2$\times$2) S covered 
Pd(100) surface can be summarized as follows:

(1) For the geometry where the molecule reaches the surface at a position 
 more than 3 \AA\  away from the sulfur atom , i.e. $d_{||} \geq 3$~\AA\ 
(e.g. at the {\bf h$_1$} site), 
the dissociation is activated and the energy barrier is about 0.1 eV in the 
entrance channel resulting from the sulfur-induced modification of 
the local electronic structure at the surface (see the elbow plots of 
{\bf b$_1$-h$_1$-b$_1$} and {\bf t$_{\rm Pd}$-h$_1$-t$_{\rm Pd}$}
geometry).

(2) For the geometry where the molecule reaches the surface at a position 
about $\sim$ 2.7 - 3.2 \AA\  away from the S atom 
(e.g. at the {\bf h$_2$} or {\bf b$_1$} sites), if the molecule is oriented so 
that the hydrogen atoms do not approach the adsorbed sulfur, the 
dissociation is also activated.  The energy barrier is in the entrance 
channel and its magnitude depends on the distance of the molecule from 
the S atoms.  For the {\bf h$_1$-b$_1$-h$_2$} geometry, 
the energy barrier is 0.16 eV, whereas the barrier height is 0.6 eV 
for the {\bf b$_2$-h$_2$-b$_2$} geometry because the latter adsorption 
position is closer to the sulfur atom.

(3) For the geometry where the molecule reaches the surface at a position 
about $\sim$ 2.7 - 3.2 \AA\  away from the sulfur atom 
(e.g. at the {\bf h$_2$} or {\bf b$_1$} sites), 
if the molecule is oriented so that the 
hydrogen atoms approach sulfur atoms, the dissociation becomes 
purely repulsive due to the direct interaction between H$_2$ 
and the S atoms (see the PES plots of 
{\bf b$_2$-h$_2$-b$_2$} and {\bf t$_{\rm Pd}$-b$_1$-t$_{\rm Pd}$} geometry). 

(4) For the geometry where the projected distance $d_{||}$ is about 
$\sim$ 1.8 - 2.3 \AA\  and if the 
molecule is oriented so that the hydrogen atoms 
do not approach the adsorbed sulfur, the dissociation behavior of the hydrogen 
molecule is activated, however, with high energy barriers ($\geq$ 1.2 eV) 
created by the strong repulsive interaction between hydrogen and sulfur
(see the PES of {\bf h$_2$-t$_{\rm Pd}$-h$_2$} geometry). 

(5) For the geometry where the projected distance $d_{||}$ is 
smaller than 1.5 \AA, the dissociation behavior of the hydrogen molecule will 
be dominated by the strong repulsive interaction between H and S atoms 
and becomes activated with a high energy barrier ($\ge$ 2.5 eV) (see PES of 
{\bf b$_2$-t$_{\rm S}$-b$_2$}) or purely repulsive if one 
of the H atoms points to the S atom. 

(6) For the geometry where the molecule is in an upright geometry with 
its molecular axis perpendicular to the surface, the PES is repulsive.

In Table~\ref{tab2} we summarize the results of the adsorption energies
and dissociation barriers at six different geometries 
at the (2$\times$2) sulfur covered surface and at the clean surface. 
For the clean Pd(100) surface, most of the hydrogen 
dissociation pathways are non-activated if the H$_2$ molecule dissociates
parallel to the surface ($\theta$=90$^\circ$), and only a few dissocation 
pathways are activated for this molecular orientation with small energy 
barriers ($\le$ 0.15 eV). It was explained by
Gross, Wilke, and Scheffler \cite{Gro95PRL} that the large initial sticking 
coefficient ($\sim$0.7) of low energy hydrogen molecules on the Pd(100) 
surface\cite{Ren89} is mainly due to the steering effect.
For the (2$\times$2) sulfur covered surface, it is interesting to see that 
all the dissociation pathways become activated
with energy barriers ranging from  0.1 eV to 2.55 eV. 
This fact explains qualtatively the reason why the sticking 
coefficient of H$_2$ is strongly reduced upon the adsorption of sulfur 
adatoms, however, a detailed understanding needs a 6-D quantum dynamical 
calculation \cite{Gro95PRL} and will be published later.\cite{Gro98} 

In order to understand the different origin of the formation of small 
energy barriers at the {\bf h$_1$} and {\bf b$_1$} sites and large energy 
barriers at {\bf t$_{\rm Pd}$}, {\bf t$_{\rm S}$} sites, we compare the 
density of states (DOS) for the H$_2$ molecule in different geometries. 
Figure 7(a) shows the DOS when the H$_2$ molecule is in
the {\bf t$_{\rm Pd}$-h$_1$-t$_{\rm Pd}$} configuration. 
The H-H distance is taken as 0.75 \AA\  and the center of mass of the H$_2$ 
molecule is 4.03 \AA\ above the topmost Pd layer such that 
there is no interaction between the hydrogen molecule and 
the sulfur covered palladium surface.
It is evident that the sulfur {\it p} orbitals strongly interact 
with the Pd {\it d} states, giving rise to a narrow peak just below 
the Pd {\it d} band edge (at $\epsilon$-$\epsilon_F$ = -4.8 eV) and 
a broad band at higher energies which has substantial DOS at the Fermi level. 
The {\it d} band at the surface Pd atoms is broadened due to the interaction 
with the S atoms. 

In Fig. 7(a), one intense peak of the DOS at the H atoms is 
found at the energy of the sulfur related bonding state at -4.8 eV. 
This degeneracy is accidental, however. There exists no broad 
distribution of states just below the Fermi level. 
This indicates that at this height above the surface, 
the $\sigma_g$ orbital of the H$_2$ molecule interacts neither 
with the sulfur related bonding state nor with the surface Pd {\it d} band. 

When the H$_2$ molecule gradually approaches the surface, the interaction 
of the hydrogen states
with the sulfur related bonding state and the surface Pd {\it d} states 
begins to build up. The intense peak of the DOS at -4.8 eV diminishs quickly 
and splits into a sharp bonding state and anti-bonding broad band just below 
the Fermi level.  In Fig. 7(b) where the H$_2$ molecule is 1.61 \AA\ above the 
topmost Pd layer, the intense DOS of the H atoms found at -4.8 eV has 
shifted. The interaction of the $\sigma_g$ orbital of the H$_2$ 
molecule with the broad band of the surface Pd {\it d} states results 
in a sharp bonding state of the H$_2$ $\sigma_g$-surface interaction 
at -7.1 eV and a broader distribution of states with substantial
weight below the Fermi level. Consequently we encounter an  occupation 
of the H$_2$-substrate antibonding states. Thus, a repulsive contribution 
to the H$_2$-surface interaction appears and gives rise to
the formation of a small energy barrier.

Figure 7(c) shows the DOS of the H$_2$ molecule in the 
{\bf b$_2$-t$_{\rm S}$-b$_2$} geometry. The H-H distance 
is taken as 0.75 \AA\  and the center of mass of the H$_2$ molecule is
3.38 \AA\ above the topmost Pd layer. It is interesting to see 
that at this height a strong interaction is already 
found between the hydrogen molecule and the adsorbed sulfur. 
The intense peak of DOS at -4.8 eV has split into a sharp bonding state 
at (-6.6 eV) and a narrow anti-bonding state (at -4.0 eV) that strongly 
indicates the direct interaction of the H$_2$ $\sigma_g$ orbital with the 
sulfur related 2 {\it p} state at -4.8 eV which leads to a large energy 
barrier. 

        Summarizing the above results and analysis, we conclude that
the non-activated dissociation at clean Pd(100) surface is inhibited 
upon the adsorption of a (2$\times$2) sulfur adlayer.  
The dissociative adsorption of hydrogen with its molecular axis parallel 
to the surface is strongly corrugated; it has a wide range of energy barriers 
(0.10 - 0.15 eV near the {\bf h$_1$} and {\bf b$_2$} sites; 
0.6 eV near the {\bf h$_2$} site; 1.4 eV near the {\bf t$_{\rm Pd}$} site; 
2.6 eV near {\bf t$_{\rm S}$}) that strongly 
depends on the projected distance $d_{||}$ between H$_{2}$ and the S atoms, 
it also depends on the molecular orientation {\bf $\phi$}.  
Still dissociative adsorption of hydrogen is an exothermic process. 
Thus the poisoning effect of sulfur adatoms for H$_2$ dissociation at 
low sulfur coverage ($\Theta_{\rm S} \leq$ 0.25) is governed by the formation 
of energy barriers and not by the blocking of adsorption sites.
To our knowledge, this is the most corrugated surface for
dissociative adsorption studied so far by {\it ab initio} calculations.
This has interesting consequences for the 
dissociation dynamics on this PES.\cite{Gro98}

\section{H$_{2}$ at the \lowercase{c(}2$\times$2) sulfur covered 
P\lowercase{d}(100) surface}
\subsection{Determination of the potential energy surface (PES)}
     The potential energy surface for the dissociative adsorption of hydrogen 
over the sulfur covered c(2$\times$2) Pd(100) surface is described within 
the same set of coordinates ($ X, Y, Z, d_{\rm H-H}, \theta, \phi$) 
as used for the (2$\times$2) surface (see Figure 8). Figure 9 shows the 
surface unit cell for the c(2$\times$2) sulfur covered Pd(100) surface.  
Inside the unit cell, we specifically analyse the dissociation at
the following sites: bridge site (denoted by {\bf b}), hollow site 
(denoted by {\bf h}), top site above the Pd atom 
(denoted by {\bf t$_{\rm Pd}$}), 
and top site above the S atom (denoted by {\bf t$_{\rm S}$}).  
To obtain a detailed information about the interaction, 
we have again calculated the elbow-plots at different sites and
different molecular orientations ($\theta, \phi$).

\subsection{Results }
   For the c(2$\times$2) S covered surface, the adsorbed S atoms form a 
4~\AA~$\times$~4~\AA\ square lattice 1.29\AA\ above the Pd layer.\cite{Wil97}
From that it follows that the projected distance $d_{||}$ of the
molecule from one sulfur adatom cannot be larger than 2.85 \mbox{\AA}.
Thus the hydrogen molecules on the Pd surface have to dissociate close
to the sulfur atoms no matter how they approach the surface.  Therefore, 
from what we have learned in the last section about the dissociation
behavior of H$_2$ over the (2$\times$2) S covered surface, we might expect 
for the c(2$\times$2) S covered surface that the 
dissociation will become purely repulsive for all 
different sites and orientations.  Indeed, this is what we 
find from our {\it ab initio} calculations. 

\subsubsection{2-D cuts through the PES at the hollow site}
  At the {\bf h} site, the projected 
distance $d_{||}$ between this site and the sulfur atom is 
2.85 \AA. However, there are four S atoms near this site so that 
a dissociating hydrogen molecule will always approach sulfur atoms.
One can thus expect that
the dissociation is purely repulsive for the hollow site.  
Figures 10(a), 10(b), and 10(c) show the PES for 
the {\bf t$_{\rm Pd}$-h-t$_{\rm Pd}$}, the {\bf b-h-b}, 
and the ($ X, Y, \theta, \phi$)= 
(0.5{\large $a$}, 0.5{\large $a$}, 0$^\circ$, 0$^\circ$) 
geometries at the hollow site, respectively.  Figure 10 shows that 
the potential energy has increased to about 1.5 eV when the hydrogen 
molecule is at the same height above the Pd surface 
as the S atoms ($Z$=1.29 \AA). As mentioned above, this 
increase is due to the strong repulsive interaction between H and S atoms.

\subsubsection{2-D cuts through the PES at top sites over Pd and S atoms}
        At the {\bf t$_{\rm Pd}$} site, the projected distance $d_{||}$ 
to the nearest sulfur atom is only 2.0 \mbox{\AA}. 
Again we can expect that the strong repulsion between the H$_2$  
and the S atoms will dominate the dissociation behavior of the molecule.  
Figures 11(a) and 11(b) show the PES for the {\bf h-t$_{\rm Pd}$-h} and 
{\bf t$_{\rm S}$-t$_{\rm Pd}$-t$_{\rm S}$} geometries 
at the {\bf t$_{\rm Pd}$} site. 
We find that the potential energy has increased to more than 2.0 eV 
even when the hydrogen molecule is at a height of $Z$=1.8 \AA\  
above the Pd atoms which is still 0.5 \AA\  
above the sulfur adlayer.

  Figure 11(c) shows the PES at the {\bf t$_{\rm S}$} site within the 
{\bf b-t$_{\rm S}$-b} geometry. In this geometry, the hydrogen molecule 
approaches the surface directly towards the S atom.
The potential energy increases
dramatically to more than 1.5 eV when the molecule is still 3 \AA\  
above the topmost Pd layer. At this geometry the PES of the c(2$\times$2) 
surface is found to be very similar to the 
PES of the (2$\times$2) surface at the 
{\bf b$_2$-t$_{\rm S}$-b$_2$} geometry. We find an energy minimum 
of 1.56 eV (at $Z$= 2.20 \AA, $d_{\rm H-H}$=1.90 \AA) as for 
the (2$\times$2) surface.  The fact that the PES of 
the (2$\times$2) and the c(2$\times$2) 
for this particular configuration are very similar shows that
the hydrogen interaction with a sulfur covered surface close to the
sulfur atoms is completely determined by the local 
bonding properties between the H and S atoms.

        Summarizing the results for the {\bf h}, {\bf t$_{\rm Pd}$}, and
{\bf t$_{\rm S}$} sites, we find that the dissociation behavior of H$_{2}$ 
at the c(2$\times$2) S covered Pd(100) surface is dominated by the
strong direct interaction between hydrogen and sulfur atoms.  
The PES for hydrogen dissociation is repulsive and endothermic due to the 
fact that for all different approach geometries the hydrogen molecules
come close to the sulfur adatoms before the dissociation
is completed.

\section{Dependence of the PES on the polar angle $\theta$ and the
azimuthal angle $\phi$ of the hydrogen molecule}
The two-dimensional cuts through the PES described in the section III and 
IV refer to the situation where the orientation of molecular axis is 
parallel or perpendicular to the surface. In reality, the molecules impinges 
on the surface with all possible orientations. At the clean Pd(100) surface, 
Wilke and Scheffler \cite{Wil96PRB} have shown that at the {\bf h-b-h} 
adsorption geometry the change of the orientation of the molecular axis 
with respect to the surface normal $\theta$ away from $\theta = 90^{\circ}$ 
implies a significant increase in the potential energy. This
increase was found to be proportional to $\cos^2 \theta$.  Eichler, Kresse, 
and Hafner \cite{Eich96} have found for the Rh(100) surface at less symmetric 
positions (halfway between bridge- and on-top position)
that the potential energy also varies similiar to the $\cos^2 \theta$ 
behavior but with the minimum of the energy shifted by 
15$^\circ$ $\sim$ 20$^\circ$ from the parallel orientation of the molecule.  

To analyze the energy variation with $\theta$ for the (2$\times$2) sulfur 
covered Pd(100) surface, we first chose one high symmetry point 
($Z$, $d_{\rm H-H}$)= (1.05 \AA, 1.0 \AA) within the 
{\bf b$_1$-h$_1$-b$_1$} geometry and calculated the potential energy 
dependence on the angle {\bf $\theta$} at this site.  
Figure 12 shows the energy variation with {\bf $\theta$} which 
is well described by $\cos^2 \theta$. To further 
analyze the energy variation with {\bf $\theta$} at less symmetric points, 
we chose  the two points ($Z$, $d_{\rm H-H}$) = (1.52\AA, 0.8\AA), 
and ($Z$, $d_{\rm H-H}$) = (1.05\AA, 1.0\AA) at the 
{\bf h$_1$-b$_1$-h$_2$} geometry and calculated their 
energy dependence on {\bf $\theta$}.  The results are shown in Fig.~13. 
The potential energy is plotted as a function of 
$\cos^2 (\theta- \theta_o)$ for values of $\theta$ between
$0^{\circ} \leq \theta \leq 180^{\circ}$. Due to the lower symmetry, the 
potential energy is no longer single-valued in this representation as in the 
high-symmetry situation of Fig.~12 with $\theta_o = 0$, but it is 
double-valued. The fact that the corresponding results fall upon one line 
indicates that the energy variation is well 
described by $\cos^2(\theta-\theta_0)$ 
with $\theta_0$ $\approx$ 2$^\circ$ - 5$^\circ$.  These findings are similar 
to the ones found for Rh(100) by Eichler and coworkers.\cite{Eich96}

The Figs. 12 and 13 demonstrate the significant increase in the potential 
energy by rotating the molecule away from {\bf $\theta$}= 90$^\circ$,
which is true at any point of the surface. This has been already indicated 
by the comparison of the PES of the hydrogen molecule approaching the 
hollow {\bf h$_1$} site in the {\bf t$_{\rm Pd}$-h$_1$-t$_{\rm Pd}$} 
geometry [see Fig. 3(b)] with the PES of the hydrogen molecule at 
the same {\bf h$_1$} site in an upright position [see Fig. 3(d)].
It is evident that rotating the molecule axis perpendicular to the 
surface increases the energy at the entrance channel and changes the 
dissociation pathway from being activated [see Fig. 3(b)] 
to being purely repulsive. 

        The effect of the azimuthal angle
{\bf $\phi$} on the hydrogen dissociation can be seen 
by comparing the PES for the {\bf t$_{\rm Pd}$-h$_1$-t$_{\rm Pd}$} 
adsorption geometry [see Fig. 3(b)] with the
{\bf b$_1$-h$_1$-b$_1$} adsorption geometry 
[see Fig. 3(a)].  For these two geometries at the {\bf h$_1$} 
site, the projected distance of the H$_2$ center of mass
from any sulfur atom is about 4 \AA\, and the two elbow plots 
are very similar at the entrance channel because the 
hydrogen atoms are still far away from any Pd or S atoms.  However, 
differences are found at the exit channel because the hydrogen atoms 
dissociate towards the Pd atoms in the {\bf t$_{\rm Pd}$-h$_1$-t$_{\rm Pd}$} 
geometry whereas they dissociate towards the bridge site for the
{\bf b$_1$-h$_1$-b$_1$} geometry.

The dependence of the PES on the azimuthal angle is even more dramatic
at adsorption sites close to the adsorbed sulfur. 
For example, if we compare the PES at the 
{\bf h$_2$} site in the {\bf b$_2$-h$_2$-b$_2$}
geometry [see Fig. 3(c)] with the {\bf b$_1$-h$_2$-b$_1$} 
geometry [see Fig. 1(c) of Ref. 10], we can see that rotating the 
molecule towards the sulfur atoms leads to a large increase in the 
potential energy and changes the dissociation from 
being activated to being purely repulsive. 
In conclusion, we expect a significant increase 
of the potential energy by rotating the molecule towards sulfur atoms if the 
projected distance $d_{||}$ of the H$_2$ center of mass
from the sulfur atoms is smaller
than 2.85\AA. 

\section{Analytical representation of the {\lowercase{\it ab initio}} \rm \bf 
PES for H$_{2}$ at S(2$\times$2)/P\lowercase{d}(100)}

In order to perform dynamical calculation on the {\it ab initio} PES,
one needs a continous analytical representation of the PES. \cite{Gro97PRB}
Details of such an representation for the PES of H$_2$ on the
clean Pd(100) surface have already been published. \cite{Gro97PRB}
For the H$_2$ dissociation on the (2$\times$2) sulfur covered
Pd(100) surface we have now also determined an analytical representation
which has been based upon the analytical PES for the clean surface.
However, due to the larger unit cell of the
S(2$\times$2)/Pd(100) surface, the analytical form has been expanded up
to the fourth-order Fourier coeffcient in the lateral directions parallel to
the surface. In addition, in the azimuthal dependence of the PES the
fourth-order term (proportional to $\cos (4\phi$)) has also been
included.

For the solution of the time-independent Schr{\"o}dinger equation 
it is necessary to transform 
the coordinates in the $Zd$-plane into reaction path coordinates
$s$ and $\rho$. \cite{Gro97PRB}
In these coordinates the function $ V (X, Y, s, \rho, \theta, \phi)$,
which  describes the potential energy surface, has the following form
in our parametrization:
\begin{equation}
V \ (X,Y,s,\rho,\theta,\phi) \ = \ V^{\rm corr} \ + \ V^{\rm rot} \ 
+ \ V^{\rm vib}
\end{equation}
with
\begin{equation}
V^{\rm corr} \ = \ \sum_{m,n = 0 }^4 \ V_{m,n}^{(1)} (s) \ \cos mGX \ \cos nGY,
\end{equation}
\begin{eqnarray}
\lefteqn{V^{\rm rot} = \sum_{m=0}^2 \ V_m^{(2)}(s) \ \frac{1}{2} 
\cos^2 \theta \  (\cos mGX + \cos mGY)} \nonumber\\
& & + \sum_{n=1}^2 \ V_n^{(3)} (s) \ \frac{1}{2} 
 \sin^2 \theta \ \cos 2 \phi \ (\cos nGX - \cos nGY) \nonumber \\
& & +  \ V^{(4)} (s) \ \frac{1}{2} 
 \sin^4 \theta \ \cos 4 \phi \ (\cos GX + \cos GY) 
\label{Vrot}
\end{eqnarray}
and
\begin{equation}
V^{\rm vib} \ = \ \frac{\mu}{2}\ \omega^2 (s)\ [\rho \ 
- \ \Delta \rho (X,Y,s)]^2 \ .
\label{V_vib}
\end{equation}
$G = 2 \pi / a$ is the length of the basis vectors of the square surface 
reciprocal lattice, $a$ is the nearest neighbor distant between  S atoms
in the (2$\times$2)S/Pd(100) surface unit cell and 
$\omega (s)$ is the vibrational frequency. 
In addition, the curvature $\kappa = \kappa (s)$ of the minimum energy path
has to be determined.
The displacement $\Delta \rho$ in the potential
term $V^{\rm vib}$ (Eq.~\ref{V_vib}) takes into account that
the location of the minimum energy path in the $Zd$-plane depends
on the cut through the six-dimensional configuration space.
$\Delta \rho$ does not influence the barrier
distribution, however, it changes the curvature of the minimum
energy paths in the $Zd$-planes. Large values of $\Delta \rho$
make the quantum dynamical calculations rather time-consuming since
they require a large number of vibrational eigenfunctions in the expansion of
the hydrogen wave function. \cite{Gro97PRB}
However, the large values of $\Delta \rho$ only occur for 
large separations of the hydrogen atoms where they do not influence the 
calculated sticking probabilities and scattering properties 
significantly. \cite{Gro97PRB} Therefore we have parametrized the displacement
properly only for values of $|\Delta \rho| \le 0.15$~{\AA}.

For the two hydrogen atoms in adjacent adsorption positions, the energetic 
cost to turn the molecule upright has not been determined by the 
{\it ab initio} calculations. This energy enters the term $V_0^{(2)}$ in
$V_{\rm rot}$ (Eq.~\ref{Vrot}). Still it is possible to estimate that 
energy cost. In the harmonic approximation we can write the potential for the
vibrations of the two hydrogen atoms perpendicular to the 
surface as
\begin{equation}
V_{\rm perp} (z_1, z_2) \ = \ \frac{m}{2} \omega_{\rm ad}^2 \ (z_1^2 + z_2^2),
\end{equation}
where $z_1$ and $z_2$ denote the $z$-coordinates of the two hydrogen
atoms, $m$ is the mass of the hydrogen atom, and $\omega_{\rm ad}$ ist
the vibrational frequency. Now we transform  $V_{\rm perp}$ to the
center-of-mass coordinate $Z = (z_1 + z_2)/2$ and the relative
coordinate $z = z_2 - z_1$. Then $V_{\rm perp}$ becomes
\begin{equation} \label{perp}
V_{\rm perp} (Z, z) \ = \ \frac{M}{2} \omega_{\rm ad}^2 Z^2 \ + \
                      \frac{\mu}{2} \omega_{\rm ad}^2 z^2 
\end{equation}
with the total mass $M = 2m$ and the reduced mass $\mu = m/2$.
The frequency $\omega_{\rm ad}$ can be determined from the 
two-dimensional cuts of the {\it ab initio} calculations.
Comparing $V_{\rm rot}$ and $V_{\rm perp}$ and using 
$z = \cos \theta \ d_{H-H}$, we obtain for
$V_0^{(2)}$ at adjacent atomic adsorption positions
\begin{equation} \label{Vperp}
V_0^{(2)} (s = s_{\rm ad}) \ = \
\frac{\mu}{2} \ \omega_{\rm ad}^2 \ d_{H({\rm ad})-H({\rm ad})}^2
\end{equation}
$V_0^{(2)}$ is therefore parametrized in such a way that it approaches the 
value of Eq.~\ref{Vperp} for the dissociated molecule on the surface.

In general the functions $V_{m,n}^{(i)}(s)$ and $\omega (s)$
are determined such that the difference to the 
{\it ab initio} calculations on the average is smaller than 50~meV.
The energies are particularly well described along the minimum energy
paths of the two-dimensional cuts of the six-dimensional configuration
space. The maximum error of the analytical representation in comparison to
the {\it ab initio} PES is of the order of 0.5~eV. This  error, however,
does only occur for large distances from the minimum energy path where
the potential energy is aleady rather high. Therefore the maximum error
has only very little influence on the dynamical calculations.
In Fig.~\ref{Fit_PES} we have plotted two cuts through the analytical
six-dimensional PES  for the  {\bf b$_1$-h$_1$-b$_1$} and the 
{\bf h$_2$-t$_{\rm Pd}$-h$_2$} geometries, respectively. 
They should be compared with the corresponding {\it ab initio} cuts 
in Fig.~\ref{fig3} and Fig.~\ref{fig5}.

\section{ CONCLUSION }
We have performed detailed calculations of the PES for the dissociative 
adsorption of hydrogen molecules 
on the (2$\times$2) and the c(2$\times$2) sulfur covered 
Pd(100) surface using density-functional theory and the full-potential 
linear-augmented plane wave method.\cite{Bla95,Koh96} The exchange correlation 
is treated in the generalized gradient approximation (GGA).\cite{Per92}
For the H$_2$ dissociation on the (2$\times$2) S covered Pd(100) surface, 
by calculating the PES over different adsorption sites we find that 
the non-activated dissociation pathways at the clean Pd(100) surface 
become activated or purely repulsive upon the adsorption of a
(2$\times$2) sulfur adlayer.  The PES is strongly corrugated. 
The minimum barrier has a height of 0.1 eV, while close to the S atoms 
the barrier towards dissociative adsorption for molecules with their axis
parallel to the surface becomes larger than 2.5 eV.
We find that the energy variation with the polar angle {\bf $\theta$} 
of the hydrogen molecule at the {\bf h$_1$} and {\bf b$_1$} 
sites is well described by 
$\cos^2(\theta-\theta_0)$ with $\theta_0 \approx 0^\circ - 5^\circ$.  
We also find that the PES strongly depends on the molecular 
azimuthal orientation {\bf $\phi$} if the projected distance of H$_2$ from the 
adsorbed sulfur is smaller than 2.85 \AA. Still the dissociative adsorption 
of hydrogen is exothermic; thus the poisoning effect 
of sulfur adatoms for H$_2$ dissociation at low sulfur coverage 
($\Theta_{\rm S} \leq$ 0.25) is governed by the formation of energy barriers, 
not by blocking of the adsorption sites.  

For the c(2$\times$2) S covered Pd(100) surface, the results of our 
calculations indicate that the interaction of H$_{2}$ with this surface 
is dominated by the strong repulsion between H$_2$  and the
sulfur atoms.  On the c(2$\times$2) S covered 
Pd(100) surface the sulfur atoms form a 4~\AA~$\times$~4~\AA\  square 
lattice above the topmost Pd layer; 
therefore, all molecules that reach the surface will 
eventually approach some adsorbed sulfur atoms before the dissociation is 
complete. Due to this fact, non-activated reaction pathways 
at the clean Pd(100) surface are completely inhibited,
and all dissociation pathways are purely repulsive.

\acknowledgements
C. M. Wei would like to acknowledge the one-year financial support from 
the National Science Council of the Republic of China.


\begin{figure}
\caption{Sketch of a coordinate system for the 
description of the hydrogen molecule above the (2$\times$2) 
sulfur covered Pd(100) surface. 
The other coordinates are the height of 
the center-of-mass of H$_2$ above the surface
$Z$, the H-H distance $d_{\rm H-H}$, 
and the angle of the molecular axis with 
the surface normal {\bf $\theta$}. }
\label{fig1}
\end{figure}
\begin{figure}
\caption{ Surface geometry of the (2$\times$2) sulfur covered Pd(100) surface 
        with two inequivalent hollow sites {\bf h$_1$}, {\bf h$_2$},
        two bridge sites {\bf b$_1$}, {\bf b$_2$}, 
        top site {\bf t$_{\rm Pd}$} (above the Pd atom),
        and {\bf t$_{\rm S}$} (above the S atom). }
\label{fig2}
\end{figure} 
\begin{figure}
\caption{ Cuts through the six-dimensional potential 
          energy surface (PES) of H$_2$ dissociation over
         (2$\times$2)S/Pd(100) at the hollow sites {\bf h$_1$} and {\bf h$_2$}:
                (a) PES for {\bf b$_1$-h$_1$-b$_1$} geometry; 
                (b) PES for {\bf t$_{\rm Pd}$-h$_1$-t$_{\rm Pd}$} geometry; 
                (c) PES for {\bf b$_2$-h$_2$-b$_2$} geometry;
                (d) PES for the molecule at the {\bf h$_1$} site 
                with the molecular axis perpendicular to the surface.
       The energy contours, given in eV per molecule,  
       are displayed as a function of the H-H distance, 
       $d_{H-H}$, and the height $Z$
       of the center-of-mass of H$_2$ above the topmost Pd layer. 
       All length scales, also in the following, are given in \mbox{\AA}.
       The geometry of each dissociation pathway is indicated above. }
\label{fig3}
\end{figure}
\begin{figure}
\caption{ Cuts through the six-dimensional potential energy surface (PES) 
          of H$_2$ dissociation over
              (2$\times$2)S/Pd(100) at the bridge site {\bf b$_1$}:
              (a) PES for {\bf h$_1$-b$_1$-h$_2$} geometry; 
              (b) PES for {\bf t$_{\rm Pd}$-b$_1$-t$_{\rm Pd}$} geometry.
         The geometry of each dissociation pathway is indicated above. }
\label{fig4}
\end{figure} 
\begin{figure}
\caption{ Cuts through the six-dimensional potential energy surface (PES) 
          of H$_2$ dissociation over  (2$\times$2)S/Pd(100) 
          at the top sites {\bf t$_{\rm Pd}$} and {\bf t$_{\rm S}$}:
          (a) PES for {\bf h$_2$-t$_{\rm Pd}$-h$_2$} geometry; (b) 
          PES for {\bf b$_2$-t$_{\rm S}$-b$_2$} geometry.
          The geometry of each dissociation pathway is indicated above.}
\label{fig5}
\end{figure}
\begin{figure}
\caption{ PES of a hydrogen molecule with $d_{\rm H-H}$= 0.76 \AA\  
       with its center of mass moving inside the (010) plane crossing 
       the {\bf h$_1$}, {\bf t$_{\rm Pd}$}, {\bf t$_{\rm S}$} sites.  
       This PES is defined as 
       ({\bf \it X},{\bf \it Y},$Z$,d,{\bf $\theta$}, 
       {\bf $\phi$})= ($t/\sqrt(2)$,$t/\sqrt(2)$,$Z$, 
       0.76\AA, 90$^\circ$, 135$^\circ$) 
       with $t$ and $Z$ as the two variables given in \AA.
       The variable $t$ is the projected distance of the 
       hydrogen center of mass from the sulfur atom in the (100) plane. 
 }
\label{fig6}
\end{figure}
\begin{figure}
\caption{ Density of states (DOS) for a H$_2$ molecule situated at 
          (a) (Z,d$_{H-H}$)= (4.03\AA, 0.75\AA) within the
                 {\bf t$_{\rm Pd}$-h$_1$-t$_{\rm Pd}$} geometry;
          (b) (Z,d$_{H-H}$)= (1.61\AA, 0.75\AA) within the
                 {\bf t$_{\rm Pd}$-h$_1$-t$_{\rm Pd}$} geometry;
          (c) (Z,d$_{H-H}$)= (3.38\AA, 0.75\AA) within the
                   {\bf b$_2$-t$_{\rm S}$-b$_2$} geometry. 
      Given is the local DOS at the H atoms, the S adatom, 
       the surface Pd atoms, and the bulk Pd atom. 
       The energies are given in eV.}
\label{fig7}
\end{figure}
\begin{figure}
\caption{ Sketch of a coordinate system for the description of 
          the hydrogen molecule above the c(2$\times$2) 
          sulfur covered Pd(100) surface. The other coordinates are
          the height of the center-of-mass of H$_2$ above the surface $Z$, 
          the H-H distance $d_{\rm H-H}$, and the angle 
          of the molecular axis with the surface normal $\theta$. }
\label{fig8}
\end{figure}
\begin{figure}
\caption{ Surface geometry of the c(2$\times$2) sulfur covered Pd(100) surface 
          with hollow site {\bf h}, bridge site {\bf b}, 
          top site {\bf t$_{\rm Pd}$} above the Pd atom, and top site
          {\bf t$_{\rm S}$} above the S atom. }
\label{fig9}
\end{figure}
\begin{figure}
\caption{ Cuts through the six-dimensional potential energy surface (PES) 
          of the H$_2$ dissociation at c(2$\times$2)S/Pd(100) 
          at the hollow site h: 
                 (a) PES for {\bf t$_{\rm Pd}$-h-t$_{\rm Pd}$} geometry;
                 (b) PES for {\bf b$_2$-h-b$_2$} geometry;
                 (c) PES for the molecule at the {\bf h} site 
                     with its axis perpendicular to the surface.
                 The geometry of each dissociation pathway
                 is indicated above. }
\label{fig10}
\end{figure}
\begin{figure}
\caption{ Cuts through the six-dimensional potential energy surface (PES) 
          of teh H$_2$ dissociation at c(2$\times$2)S/Pd(100) at the 
          top sites {\bf t$_{\rm Pd}$} and {\bf t$_{\rm S}$}:
               (a) PES for {\bf h-t$_{\rm Pd}$-h} geometry; 
               (b) PES for {\bf t$_{\rm S}$-t$_{\rm S}$-t$_{\rm S}$} geometry;
               (c) PES for {\bf b-t$_{\rm S}$-b} geometry.
        The geometry of each dissociation pathway is indicated above. }
\label{fig11}
\end{figure}
\begin{figure}
\caption { Dependence of the potential energy of the H$_2$ molecule 
           at the (2$\times$2) sulfur covered Pd(100) surface on 
           the angle {\bf $\theta$} of its 
           axis with the surface normal. 
           The potential energy is plotted as a function of $\cos^2 \theta$.
           The configuration for 
           {\bf $\theta$} = 90$^\circ$ corresponds to the
           {\bf b$_1$-h$_1$-b$_1$} geometry with 
           ($d_{H-H}$, $Z$) = (1.0\AA, 1.05\AA). }
\label{fig12}
\end{figure}
\begin{figure}
\caption{ Dependence of the potential energy of the H$_2$ molecule 
           at the (2$\times$2) sulfur covered Pd(100) surface on 
           the angle {\bf $\theta$} of its 
          axis with the surface normal shifted by $\theta_o$. 
          The potential energy is plotted as a function of 
          $\cos^2 (\theta- \theta_o)$ for values of $\theta$ between
          $0^{\circ} \leq \theta \leq 180^{\circ}$, i.e. the potential
          energy is double-valued in this representation. 
          The configuration for {\bf $\theta$}= 90$^\circ$ corresponds to 
          the {\bf h$_1$-b$_2$-h$_2$} geometry: 
         (a) ($d_{H-H}$, $Z$, 
             $\theta_o$) = (1.52\AA, 0.8\AA, 4.5$^\circ$);
         (b) ($d_{H-H}$, $Z$, 
             $\theta_o$) = (1.0\AA, 1.05\AA, 2.2$^\circ$). }
\label{fig13}
\end{figure}
\begin{figure} 
\caption{Two cuts through the analytical representation of the
six-dimensional {\it ab initio} PES  for the  {\bf b$_1$-h$_1$-b$_1$} and the 
{\bf h$_2$-t$_{\rm Pd}$-h$_2$} geometries, respectively.}
\label{Fit_PES}
\end{figure}

\clearpage
\begin{table}
\widetext
\normalsize
\begin{center}
\begin{tabular}{|c|c|c|c|}
geometry &\ ({\bf \it Z, d})\ &\ 
3 Pd slab, $E_{\rm cut}$= 11 Ry\ &\ 5 Pd slab, $E_{\rm cut}$= 14 Ry\ \\ \hline
{\bf b$_1$-h$_1$-b$_1$} &(0.61 \AA,0.97 \AA) & 0.28 eV &  0.24 eV \\ \hline
{\bf h$_1$-b$_1$-h$_2$} &(0.08 \AA,2.85 \AA) &-0.79 eV & -0.72 eV \\ 
\cline{2-4}   &(0.97 \AA,1.14 \AA) & 0.13 eV &  0.08 eV \\ \hline
{\bf t$_{\rm Pd}$-h$_1$-t$_{\rm Pd}$} &(1.05 \AA,0.94 \AA) &-0.02 eV 
& -0.01 eV \\ \hline
{\bf h$_2$-t$_{\rm Pd}$-h$_2$} &(1.45 \AA,1.29 \AA) & 1.30 eV & 1.31 eV \\ 
\end{tabular}                        
\end{center}
\caption{ Dependence of the potential energy on the thickness of the substrate 
          Pd slab  and the cut-off energy at different H$_2$ geometries. 
           The energies are given per H$_2$ molecule.}
\label{tab1}
\end{table}

\begin{table}
\widetext
\normalsize
\begin{center}
\begin{tabular}{|c|c|c|} 
adsorption site and geometry & clean surface ($\Theta_{\rm S}$=0) & 
(2$\times$2) surface ($\Theta_{\rm S}$=0.25) \\ \hline
{\bf h$_1$} ($\theta$,$\phi$)=(90$^o$,90$^o$) &   
& activated E$_{\rm b}$= 0.10 eV \\
\cline{1-1} \cline{3-3}
{\bf h$_2$} ($\theta$,$\phi$)=(90$^o$,90$^o$) & non-activated, 
E$_{\rm ad}$= 0.46 eV & activated E$_{\rm b}$= 0.60 eV\\ 
\cline{1-1} \cline{3-3}
{\bf h$_2$} ($\theta$,$\phi$)=(90$^o$,\ 0$^o$) &  & repulsive ($\ge$ 0.75 eV) 
\\ \hline
{\bf h$_1$} ($\theta$,$\phi$)=(90$^o$,135$^o$) & 
non-activated, E$_{\rm ad}$= 0.23 eV & activated E$_{\rm b}$= 0.13 eV \\ \hline
{\bf h$_1$} ($\theta$,$\phi$)=(\ 0$^o$,\ 0$^o$) & 
no dissociation & repulsive ($\ge$ 0.25 eV)\\ \hline
{\bf b$_1$} ($\theta$,$\phi$)=(90$^o$,90$^o$) 
& non-activated, E$_{\rm ad}$= 1.22 eV & activated E$_{\rm b}$= 0.15 eV \\ 
\hline
{\bf b$_2$} ($\theta$,$\phi$)=(90$^o$,\ 0$^o$) & no dissociation & 
repulsive ($\ge$ 0.25 eV) \\ \hline
{\bf t$_{\rm Pd}$} ($\theta$,$\phi$)=(90$^o$,135$^o$) & 
activated, E$_{\rm b}$= 0.16 eV & activated E$_{\rm b}$= 1.28 eV\\ \hline
{\bf t$_{\rm S}$} ($\theta$,$\phi$)=(90$^o$,\ 0$^o$) & ----- & 
activated E$_{\rm b}$= 2.55 eV\\ 
\end{tabular}                        
\end{center}
\caption{ Summary of the results for adsorption energies and barrier heights
          for the H$_2$ dissociation over the clean Pd(100) surface
          and the (2$\times$2) sulfur covered Pd(100) surface at 
          different adsorption sites and geometries. The energies
          are given per H$_2$ molecule.}
\label{tab2}
\end{table}
\clearpage
\clearpage
\newcounter{cms}
\setlength{\unitlength}{1mm}
\thicklines
\large
\pagestyle{empty}
\begin{picture}(120.0, 80.0)
\put(0,-180){
\psfig{figure=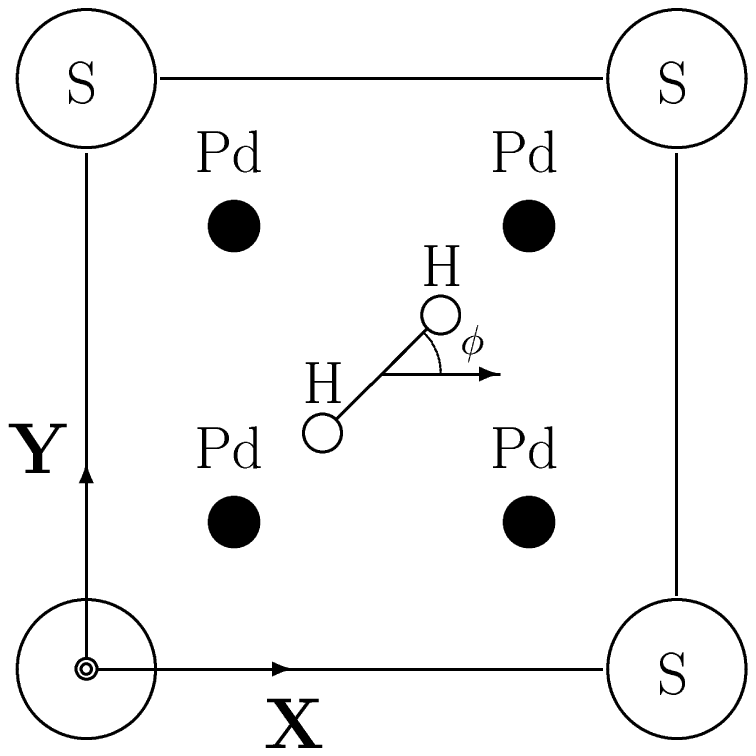,height=30cm,angle=0} }

\put(  35.00, -10.00){\huge{Fig. 1}}

\put(  35.00,-130.00){\huge{Fig. 2}}

\put(0,-300){
\psfig{figure=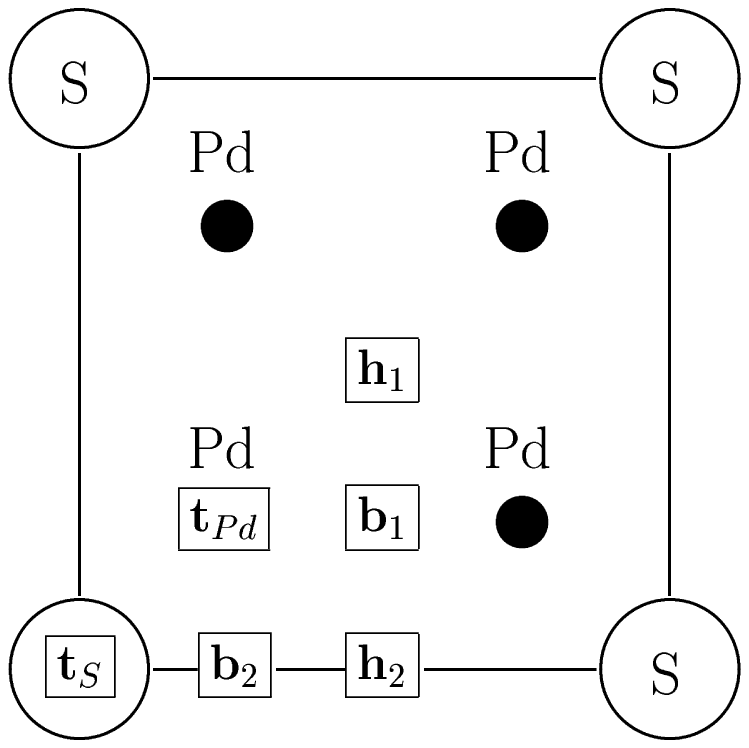,height=30cm,angle=0} }

\end{picture}

\clearpage
\begin{picture}(100,100)(+0,120)
\put(-0,130){
\psfig{figure=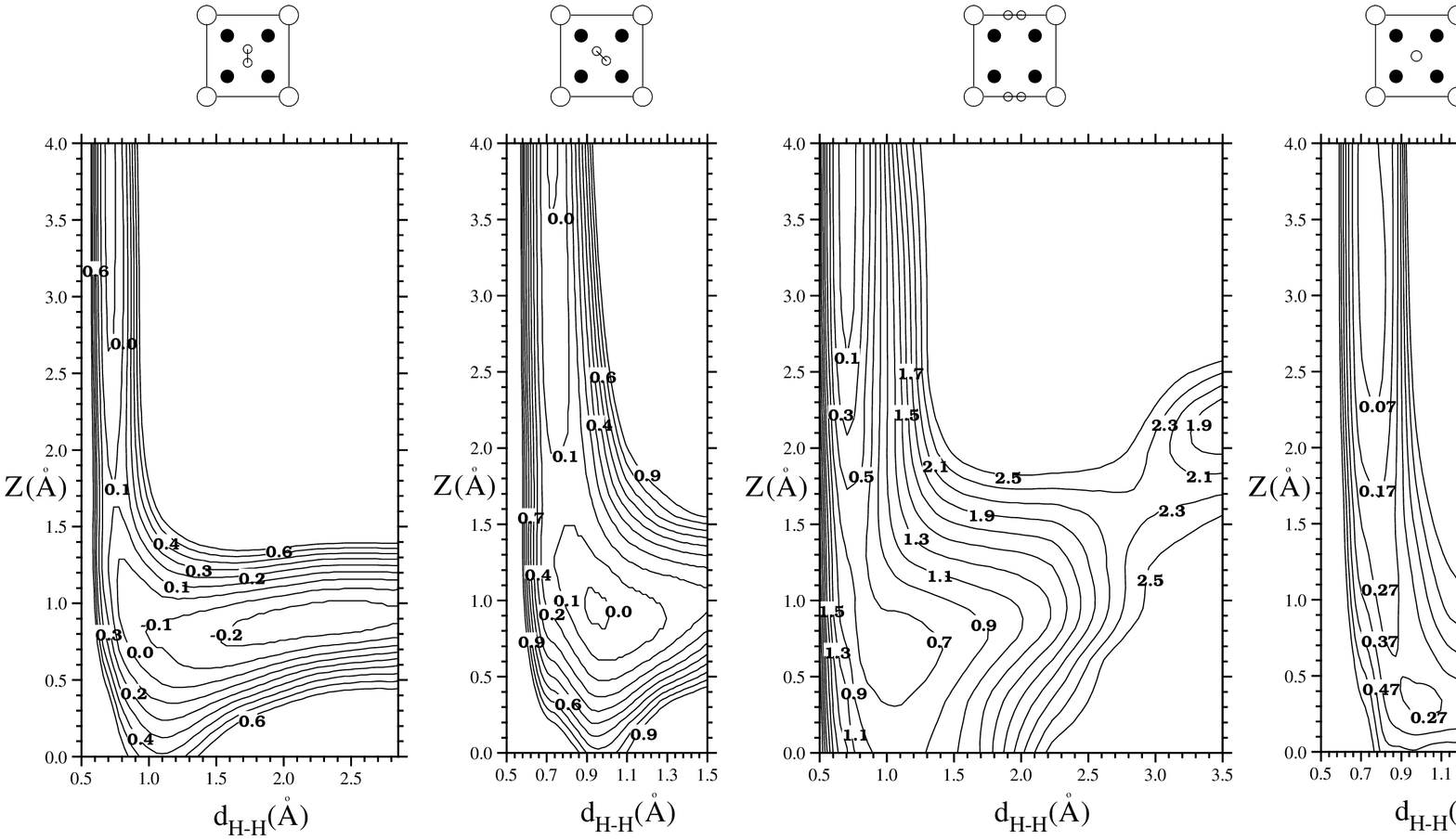,height=10cm,angle=-90} }
\put( 40,145.00){\huge{Fig. 3}}
\put(-0,40){
\psfig{figure=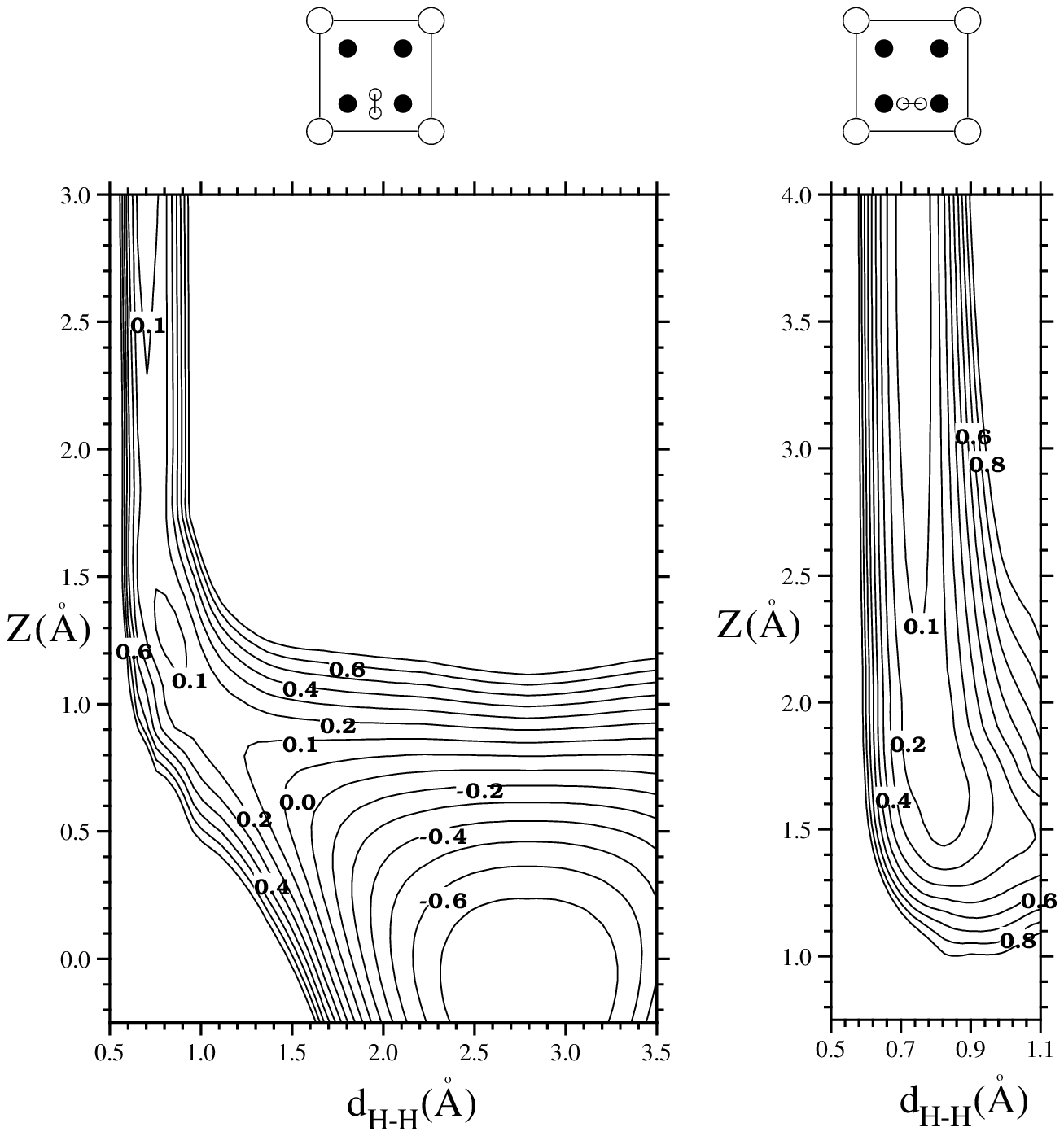,height=10cm,angle=-90} }
\put( 40,55.00){\huge{Fig. 4}}
\put(-0,-50){
\psfig{figure=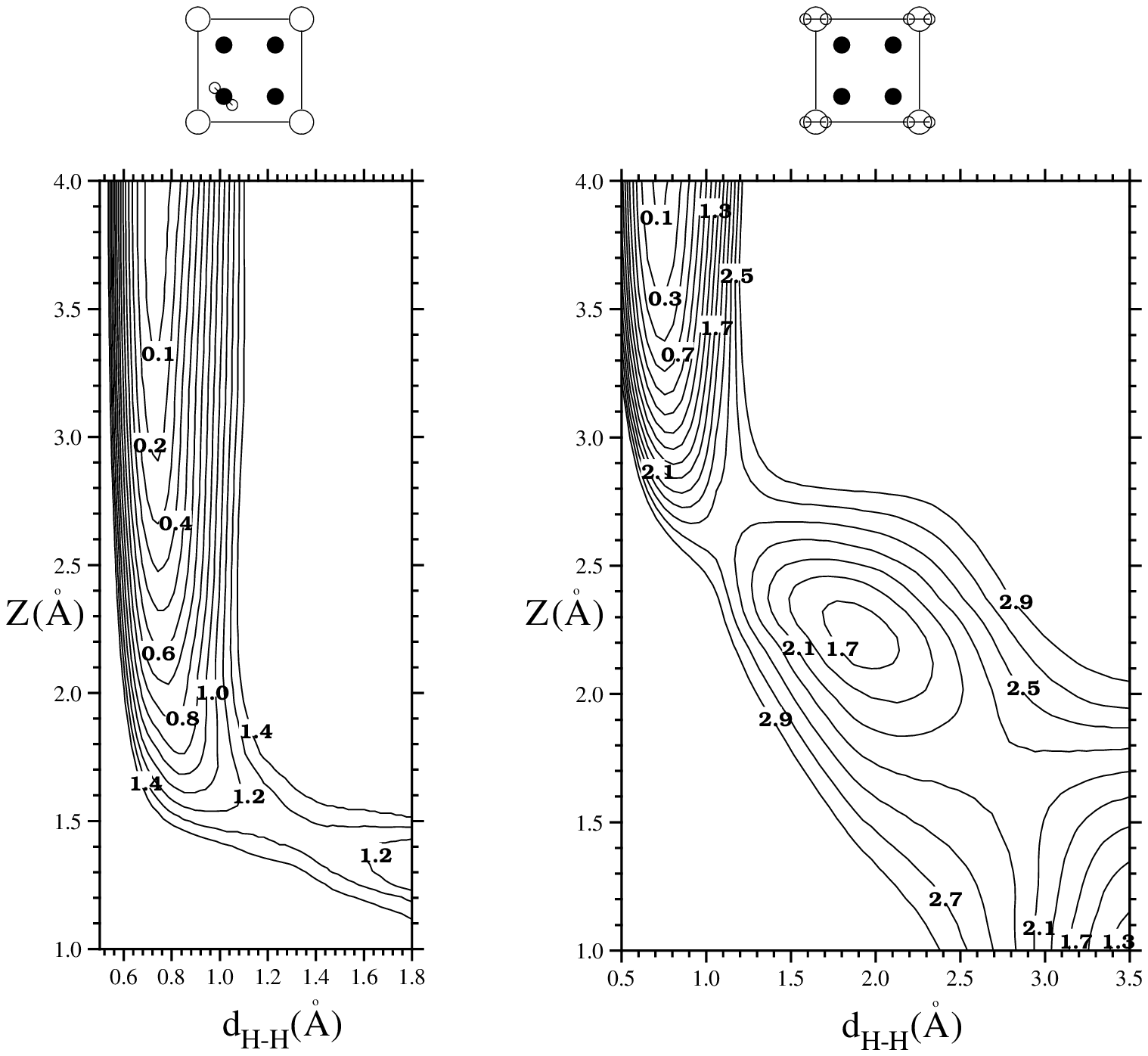,height=10cm,angle=-90} }
\put( 40,-35.00){\huge{Fig. 5}}
\end{picture}


\clearpage
\begin{picture}(100,100)(+0,120)
\put(-50,50){
\psfig{figure=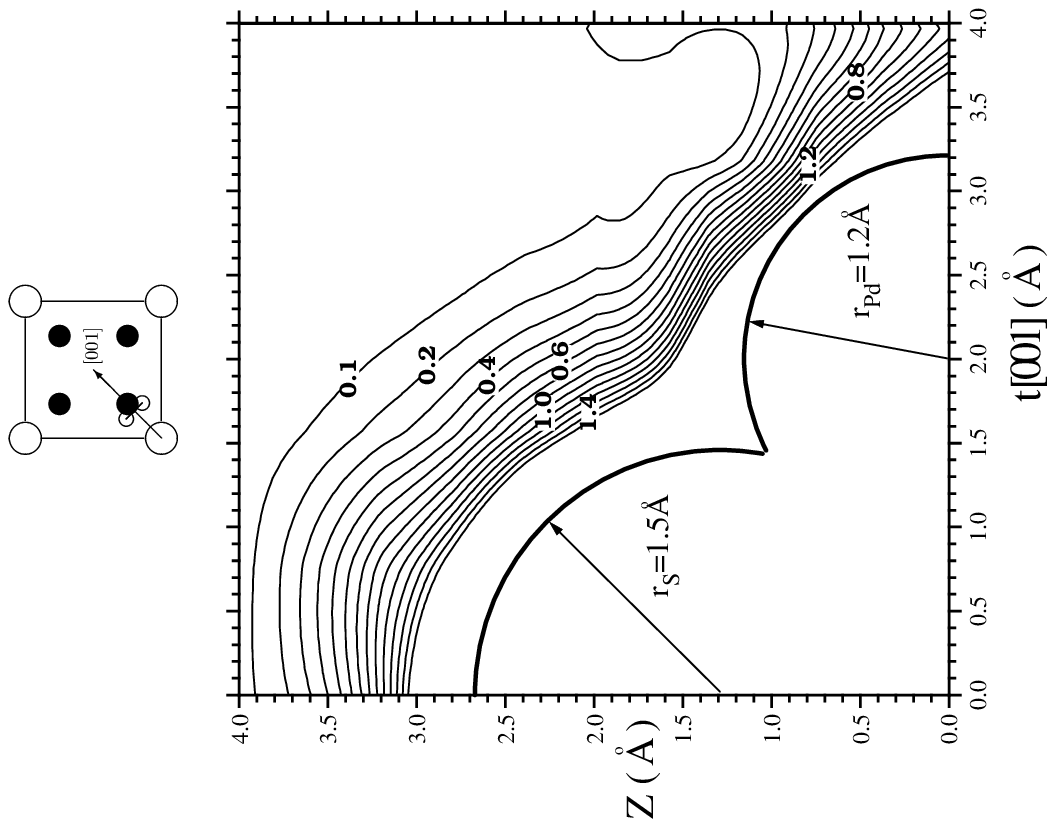,height=18cm,angle=-90} }
\put( 40,110.00){\huge{Fig. 6}}
\put(-50,-50){
\psfig{figure=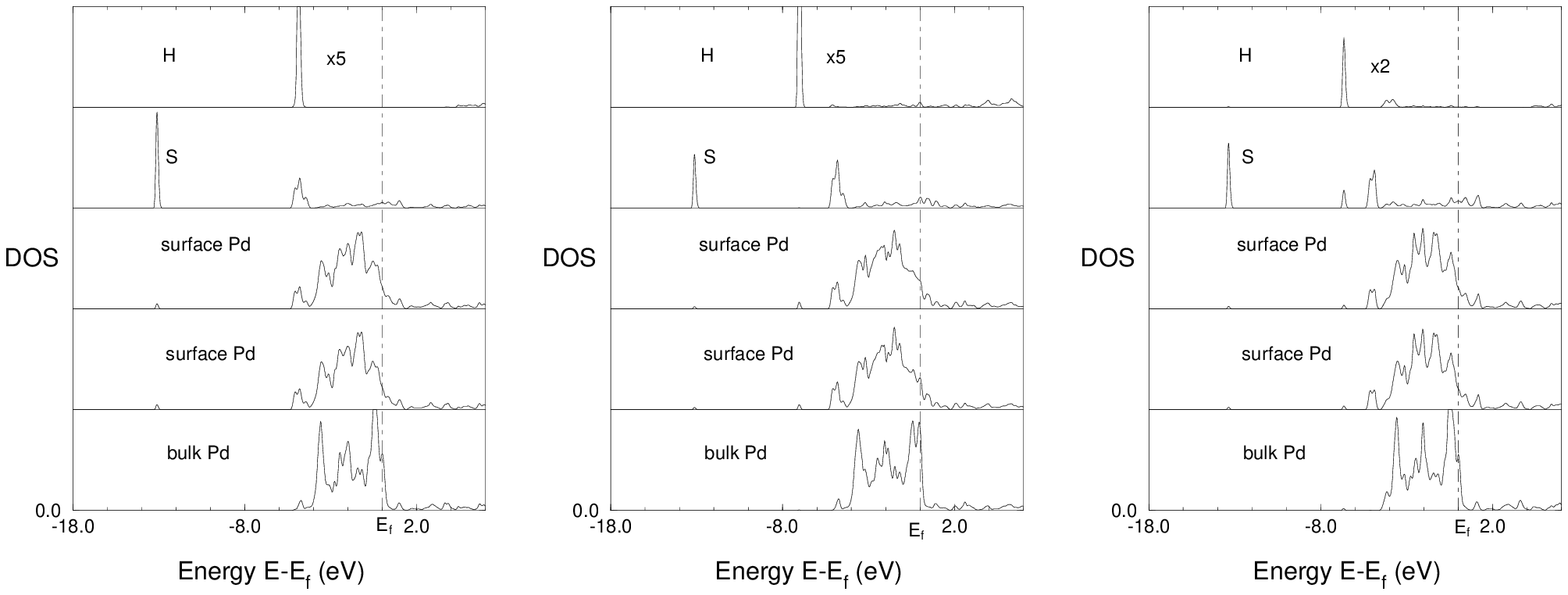,height=18cm,angle=-90} }
\put( 40,-5.00){\huge{Fig. 7}}
\end{picture}

\clearpage
\setlength{\unitlength}{1mm}
\thicklines
\large
\begin{picture}(120.0, 80.0)
\put(0,-180){
\psfig{figure=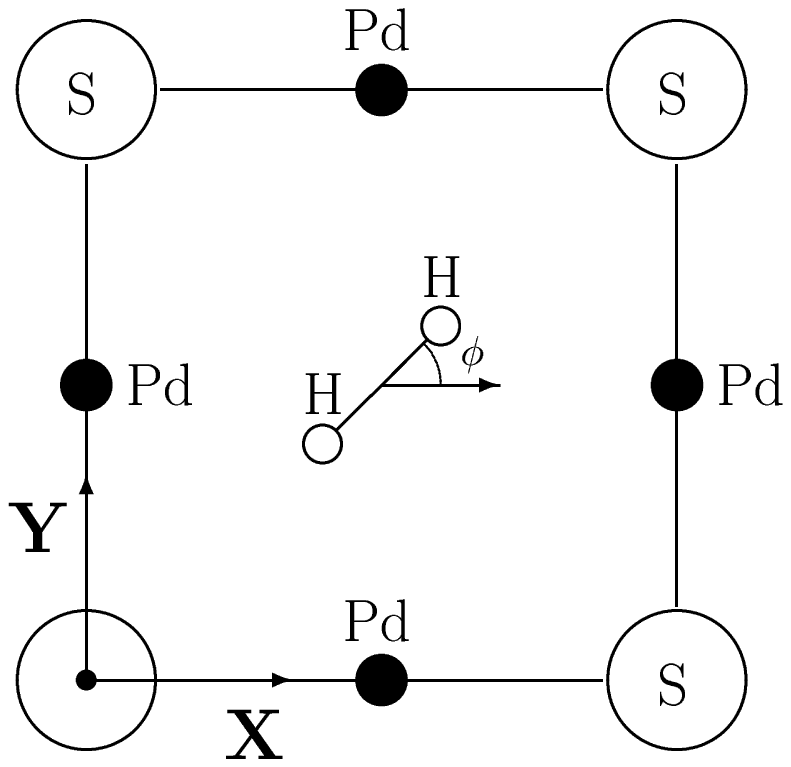,height=30cm,angle=0} }

\put(  35.00, -10.00){\huge{Fig. 8}}

\put(  35.00,-130.00){\huge{Fig. 9}}

\put(0,-300){
\psfig{figure=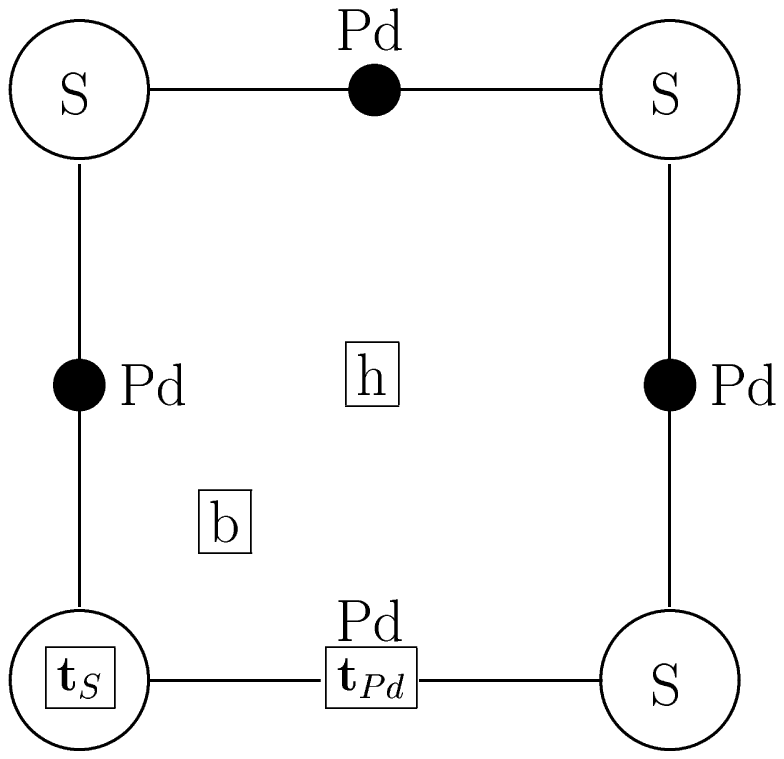,height=30cm,angle=0} }

\end{picture}
\clearpage
\begin{picture}(100,100)(+0,120)
\put(-0,130){
\psfig{figure=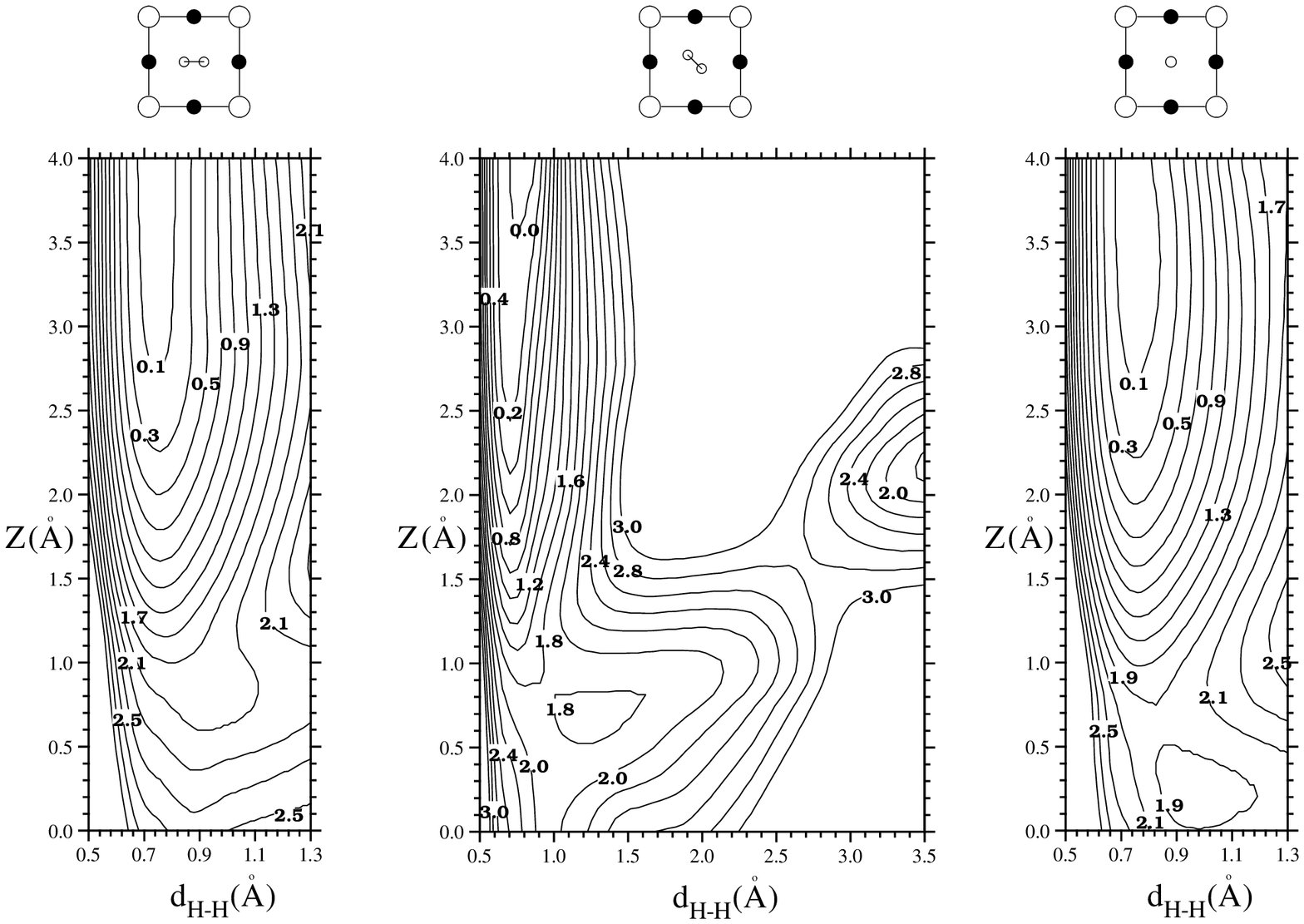,height=10cm,angle=-90} }
\put( 40,145.00){\huge{Fig. 10}}
\put(-0,40){
\psfig{figure=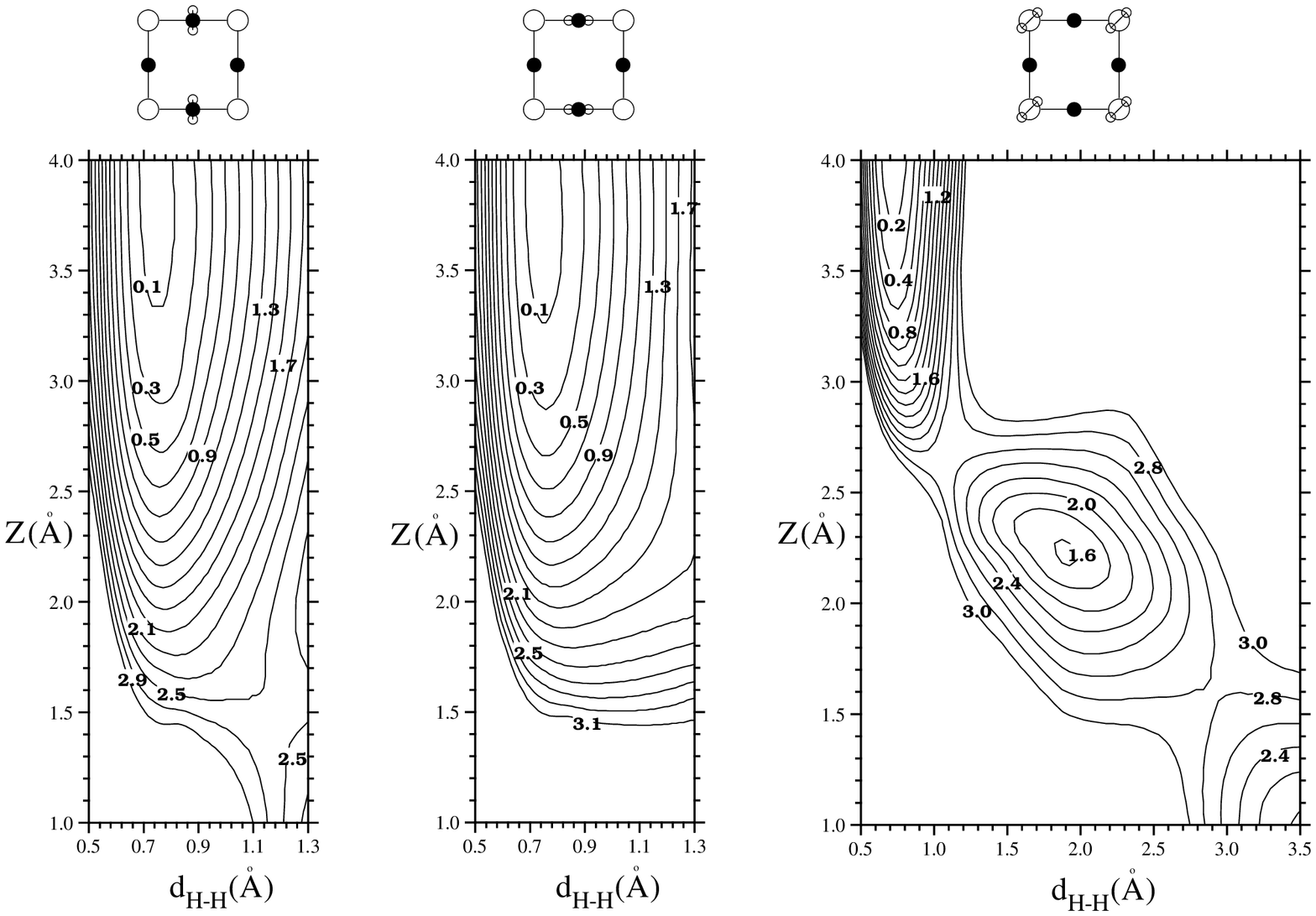,height=10cm,angle=-90} }
\put( 40,55.00){\huge{Fig. 11}}
\end{picture}


\clearpage
\begin{picture}(100,100)(+0,120)
\put(-00,120){
\psfig{figure=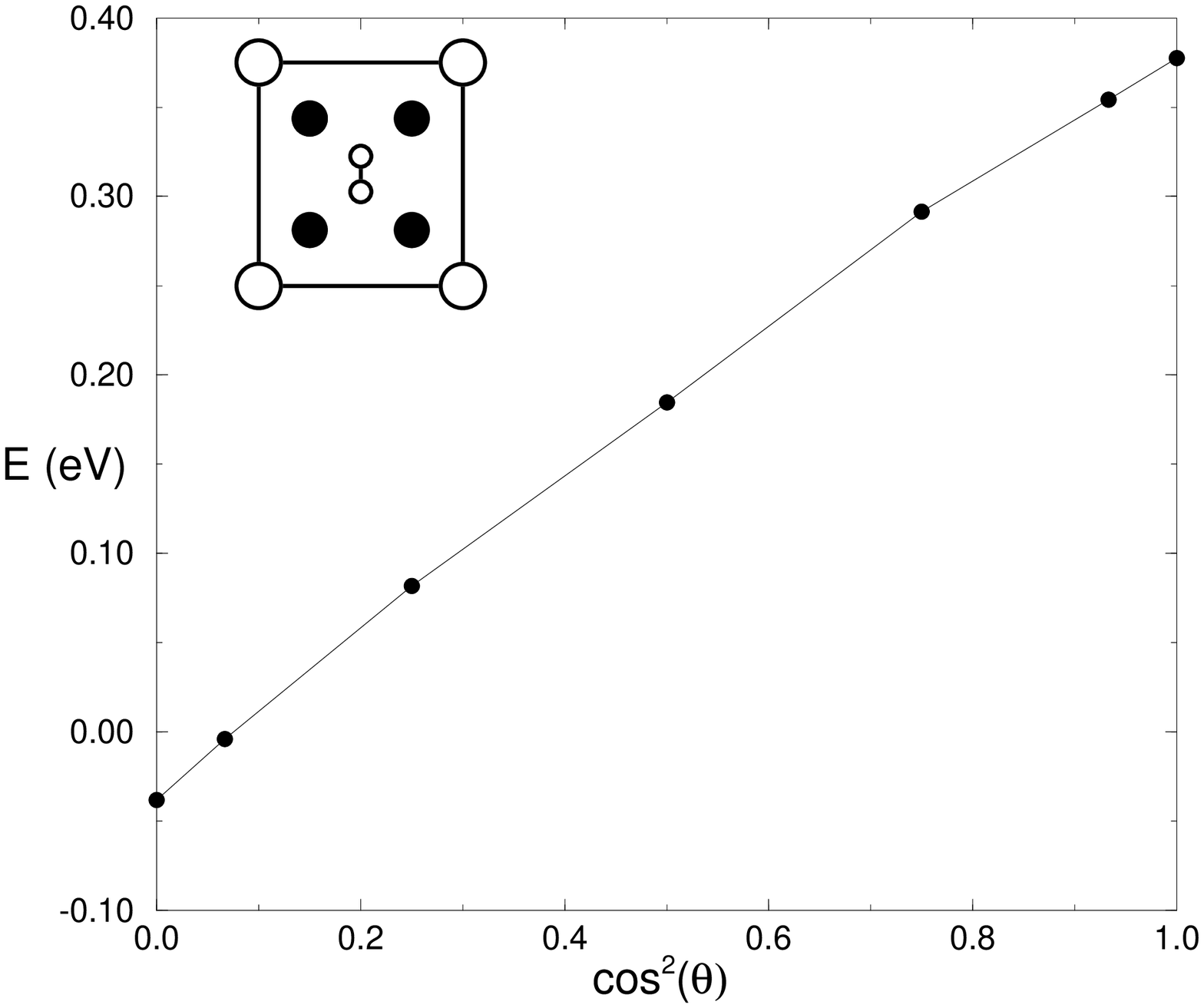,height=10cm,angle=-90} }
\put( 40,110.00){\huge{Fig. 12}}
\put(-00,+00){
\psfig{figure=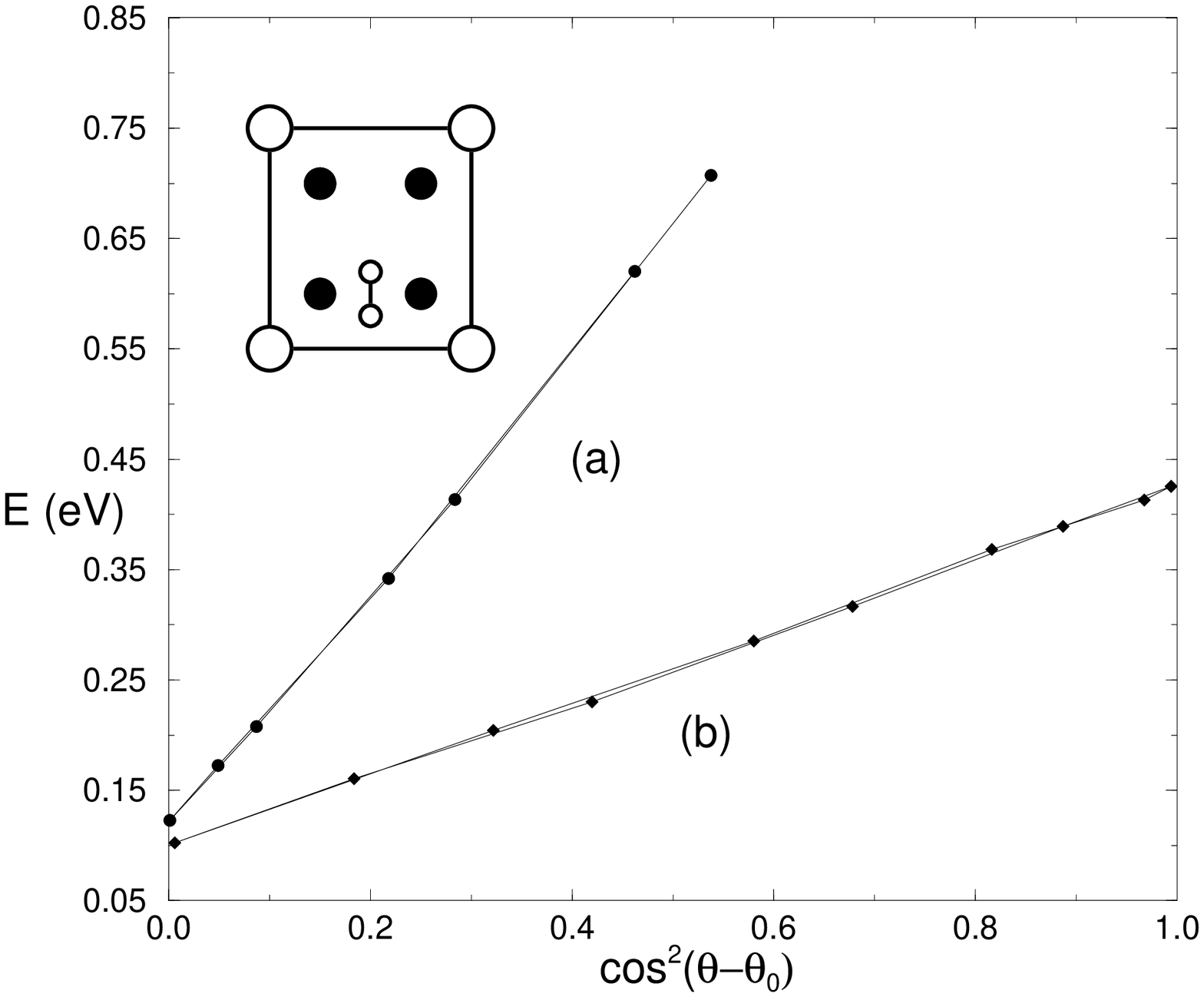,height=10cm,angle=-90} }
\put( 40,-20.00){\huge{Fig. 13}}
\end{picture}
\clearpage
\begin{picture}(100,100)(+0,120)
\put(-180,0){
\psfig{figure=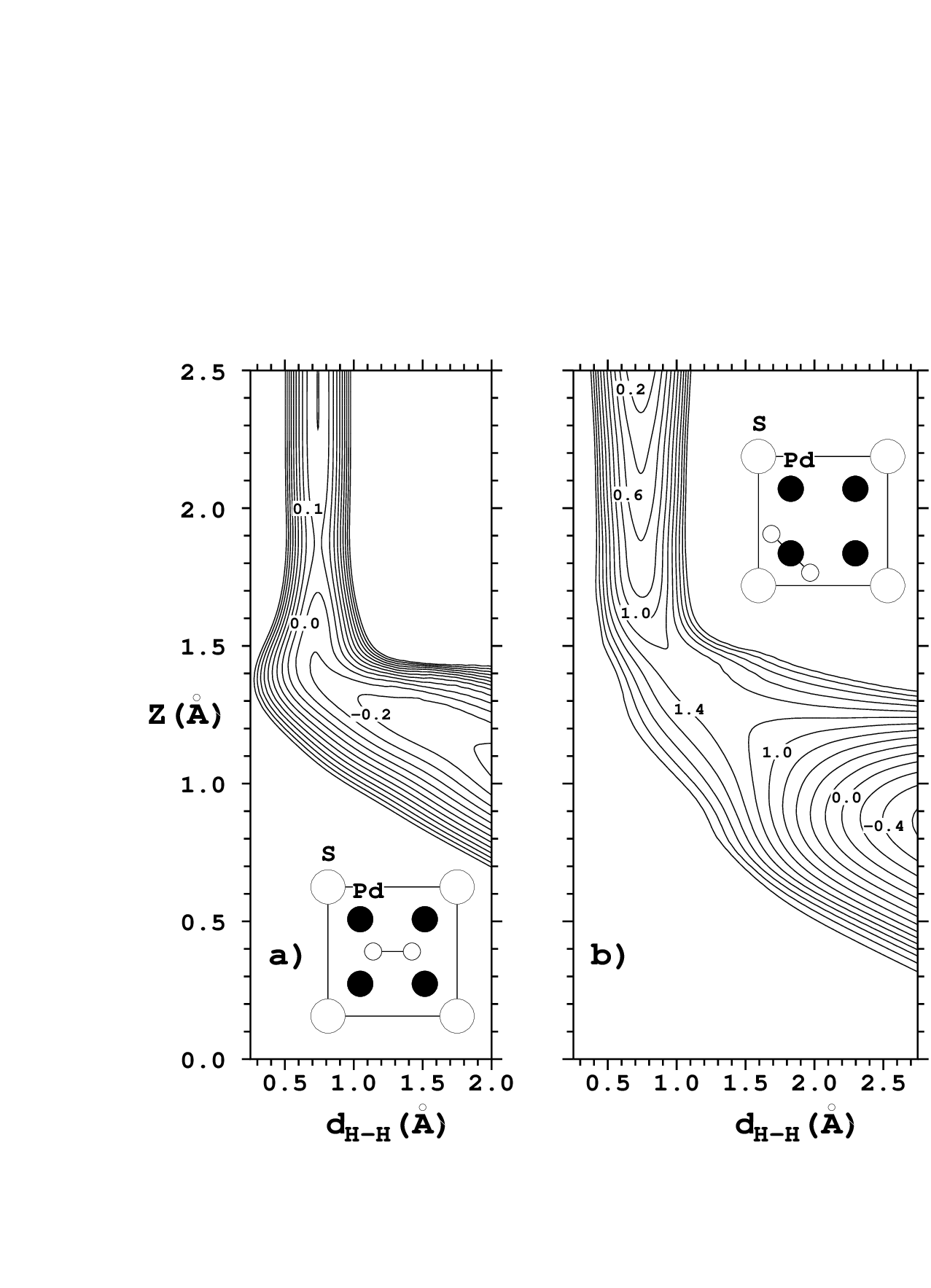,
           height=8cm,angle=00} }
\put( 40,-5.00){\huge{Fig. 14}}
\end{picture}

\end{document}